\documentclass[reprint,amsmath,amssymb,superscriptaddress,prb]{revtex4-2}

\usepackage{graphicx}
\usepackage{dcolumn}
\usepackage{bm}
\usepackage{color}
\usepackage{amsmath}
\usepackage{booktabs} 
\usepackage{array} 
\usepackage{makecell}
\usepackage{colortbl}
\usepackage{xcolor} 

\begin{document}

\title{Low-energy proton impact dynamics on hydrocarbons: Dependence on kinetic energy and incident site}

\author{Misa Viveiros}
\affiliation{Department of Physics and Astronomy, Vanderbilt University, Nashville, TN 37235, USA}

\author{Roy Lau}
\affiliation{Department of Physics and Astronomy, Vanderbilt University, Nashville, TN 37235, USA}

\author{Samuel S. Taylor}
\affiliation{Department of Physics and Astronomy, Vanderbilt University, Nashville, TN 37235, USA}
\affiliation{Pritzker School of Molecular Engineering, University of Chicago, Chicago, IL 60637, USA}

\author{Patrick Barron}
\affiliation{Department of Physics and Astronomy, Vanderbilt University, Nashville, TN 37235, USA}

\author{Attila Czirj\'ak}
\affiliation{ELI ALPS, ELI-HU Non-Profit Ltd, Wolfgang Sandner utca 3., 6728 Szeged, Hungary}
\affiliation{Department of Theoretical Physics, University of Szeged, Tisza L. krt. 84-86, 6720 Szeged, Hungary}

\author{Cody Covington}
\affiliation{Department of Chemistry, Austin Peay State University,
Clarksville, TN 37044, USA}

\author{K\'alm\'an Varga}
\email{kalman.varga@vanderbilt.edu}
\affiliation{Department of Physics and Astronomy, Vanderbilt University, Nashville, TN 37235, USA}

\begin{abstract}
The dynamics of low-energy proton collisions with hydrocarbon
molecules are investigated using real-time time-dependent density
functional theory (TDDFT). Through systematic variation of proton
kinetic energy and impact site on the molecular surface, the resulting
scattering, proton capture, and bond dissociation pathways are
analyzed. The simulations reveal a strong dependence of reaction
outcomes on both incident energy and collision geometry, with the
interplay between electronic and nuclear degrees of freedom
highlighted as governing molecular fragmentation and reaction
mechanisms. These findings provide insight into the fundamental
processes underlying proton–hydrocarbon interactions relevant to
radiation chemistry, ion-beam processing, astrochemical environments,
and Coulomb explosion.
\end{abstract}

\maketitle

\section{Introduction}
Ion–molecule collisions are fundamental to a broad range of physical, chemical, and biological processes, spanning from radiation damage in organic matter~\cite{ijms25126368,PhysRevLett.130.118401,KC2022,10.1093/mam/ozae044.902} and ion-beam cancer therapy~\cite{KRAFT20001,cancers7010353,Li2024,GRAEFF2023104046} to ion–beam–induced material modification~\cite{Huang2024,SHABI2025105872,LIU2025163442,Krasheninnikov2007,PhysRevB.85.235435,PhysRevLett.114.063004} and astrochemical reaction pathways~\cite{https://doi.org/10.1002/2014JA020208,refId0,Cui2024}. Among these diverse applications, particular attention has been devoted to the interaction of protons and ions with hydrocarbons and other biologically relevant molecules, owing to the ubiquity of C–H containing species in planetary atmospheres, the interstellar medium, and organic materials~\cite{Lüdde_2019,PhysRevA.75.042711,PhysRevLett.103.213201,Abdurakhmanov_2018,Leung2019,Seraide2017,
PhysRevA.95.052701,Wang_2022}. In such systems, both the kinetic energy of the incident proton and the collision geometry critically determine whether the interaction results in elastic scattering, chemical bond formation, or molecular fragmentation~\cite{Wang_2022}.

Research on proton-molecule collisions remains limited, with most
existing calculations focusing primarily on the keV energy range
\cite{10.1063/1.59216,J_M_Sanders_2003,BAUMGART19841,atoms8030059,
Lüdde2019,PhysRevLett.106.163001,10.1063/1.5088690,PhysRevA.95.042517}.
Time-dependent density functional theory  has been employed to
model proton-DNA collisions at 4 keV energies, with the primary
objective of elucidating DNA base pair dissociation mechanisms as a
function of proton impact locations \cite{Seraide2017}.
Ab initio molecular dynamics simulations were utilized to investigate
proton collisions with deoxyribose, where 5-7 eV protons generated
from Coulomb explosions of doubly ionized water molecules subsequently
impact deoxyribose, resulting in ring opening 
\cite{herve_du_penhoat_proton_2018}. This study was
constrained to the adiabatic regime.
The influence of collision sites on radiation dynamics in cytosine
following proton bombardment in the 150-1000 eV energy range has been
examined, along with investigations of proton-water scattering
processes \cite{Wang_2022}. Additional studies have explored 
collisions between oxygen molecules (O$_2$) with kinetic energies of 4, 6, or 10 eV and stationary
target molecules (Mg$_2$, SiH$_4$, or CH$_4$) to understand light emission
phenomena during combustion processes
\cite{miyamoto_molecular-scale_2019}. Regarding proton-hydrocarbon
collisions specifically, computational studies are particularly
scarce. The proton collision dynamics on CH$_4$ has been investigated 
\cite{gao_theoretical_2014,PhysRevA.95.042517} using 30 eV
protons and the TDDFT calculations demonstrated good agreement with experimental
observations for fragment distribution patterns.

Low-energy proton-molecule scattering exhibits fundamentally different
behavior compared to high-energy collisions due to the critical role
of specific interaction sites. During high-energy encounters, rapidly
moving protons traverse the molecular structure while transferring
only a small portion of their energy to electronic excitations, which
subsequently cascade into nuclear motion
\cite{PhysRevB.85.235435,PhysRevA.95.052701}. Conversely, low-energy
collisions result in dramatic alterations to both the proton's kinetic
energy and trajectory, making the reaction outcome highly sensitive to
the precise impact location.
These low-energy systems are characterized by complex interactions
arising from multiple scattering centers, anisotropic molecular
potential surfaces, and varied bonding environments. This complexity
generates a diverse array of possible reaction channels, encompassing
elastic scattering events, proton capture processes, and atomic
abstraction reactions.

Low-energy proton collisions are also highly relevant in the context of hydrocarbon 
Coulomb explosion. Under intense laser–molecule interactions, hydrocarbons undergo rapid 
ionization, leading to the production of a wide range of ionic fragments. The lightest and 
most abundant of these fragments are 
protons~\cite{PhysRevA.111.013109,Last2002,Cornaggia_1992,Mogyorosi2025,PhysRevLett.92.063001,D3CP02337K,PhysRevA.111.033109,PhysRevA.82.043433,Livshits2006}, which are typically expelled with kinetic energies ranging from a few to several tens of electronvolts~\cite{PhysRevLett.106.163001,PhysRevLett.92.063001}. These protons can subsequently collide with nearby neutral or ionized hydrocarbons in the gas phase, initiating secondary processes such as additional fragmentation, chemical rearrangements, and the formation of new molecular species—processes that occur alongside the primary light–matter interaction. Previous theoretical studies of Coulomb explosion have predominantly focused on the dynamics of isolated molecules, thereby neglecting the role of fragment–molecule collisions~\cite{PhysRevA.111.013109,Mogyorosi2025,PhysRevA.111.033109,Livshits2006}. By investigating proton–hydrocarbon collisions in this low-energy regime, we seek to examine these secondary pathways and assess their potential contribution to the overall reaction dynamics.

Time-dependent density functional theory \cite{runge1984density,ullrich} 
provides a powerful first-principles framework for simulating nonequilibrium processes 
in real time, as it captures both the electronic response and the coupled nuclear dynamics
\cite{kononov_reproducibility_2024,hekele_all-electron_2021,herring_recent_2023,xu_real-time_2024,
wang_efficient_2015,rossi_kohnsham_2017,noda_salmon_2019,andrade_time-dependent_2012,
andrade_time-dependent_2012,fuks_time-dependent_2016,dar_reformulation_2024,doi:10.1021/acs.chemrev.0c00223}.
By accounting for nonadiabatic effects, 
TDDFT has been successfully applied to describe Coulomb 
explosion~\cite{PhysRevA.111.013109,Mogyorosi2025,PhysRevA.111.033109,Livshits2006} as well as ion–molecule collisions~\cite{PhysRevB.85.235435,PhysRevA.95.052701,Wang_2022,Seraide2017,PhysRevA.95.042517,PhysRevLett.114.063004}.

In this work, we employ real-time TDDFT to investigate proton collisions with representative hydrocarbons\textemdash C\textsubscript{2}H\textsubscript{2}, C\textsubscript{3}H\textsubscript{8}, and C\textsubscript{4}H\textsubscript{10}\textemdash across a range of proton kinetic energies (0.52–87.58~eV) and incident sites. 
This energy window lies firmly within the low-energy regime, and it
corresponds closely to the kinetic energies of protons produced in
Coulomb explosion ~\cite{PhysRevLett.106.163001,PhysRevLett.92.063001}. By systematically 
varying both the projectile energy and the collision geometry, we identify distinct 
regimes of scattering, proton capture, and atom abstraction, and quantify the dependence 
of these processes on the collision parameters.

Our results provide new insight into proton–hydrocarbon interactions on femtosecond timescales, revealing systematic trends that can guide future experimental studies and support the development of more comprehensive models of radiation-induced chemistry in complex molecular systems. They also emphasize the important role of many-body effects in governing Coulomb explosion dynamics.

\section{Computational Method}
\label{sec:computational-method}

The simulations were performed using TDDFT for modeling the electron dynamics on a 
real-space grid with real-time propagation \cite{Varga_Driscoll_2011a}, 
with the Kohn-Sham (KS) Hamiltonian of the following form
\begin{equation}
\begin{split}
\hat{H}_{\text{KS}}(t) = -\frac{\hbar^2}{2m} \nabla^2 + V_{\text{ion}}(\mathbf{r},t) + 
V_{\text{H}}[\rho](\mathbf{r},t) \\
+ V_{\text{XC}}[\rho](\mathbf{r},t).
\end{split}
\label{eq:hamiltonian}
\end{equation}
Here, $\rho$ is the electron density, defined as the sum of the densities of all occupied orbitals:
\begin{equation}
\rho(\mathbf{r},t) = \sum_{k=1}^{\infty} f_k |\psi_k(\mathbf{r},t)|^2,
\end{equation}
where $f_k$ is the occupation number of the orbital $\psi_k$, which can take values 0, 1, or 2. 
Additionally, $f_k$ must satisfy the constraint $\smash{\sum_{k=1}^{\infty}} f_k = N$, where $N$ is the total number of valence electrons in the system.

$V_{ion}$ in eq.~\ref{eq:hamiltonian} is the external potential due to the ions, represented by employing norm-conserving pseudopotentials centered at each ion as given by Troullier and Martins~\cite{PhysRevB.43.1993}. $V_{H}$ is the Hartree potential, defined as
\begin{equation}
V_H(\mathbf{r}, t) = \int \frac{\rho(\mathbf{r}', t)}{|\mathbf{r} - \mathbf{r}'|} \, d\mathbf{r}',
\end{equation}
and accounts for the electrostatic Coulomb interactions between electrons. The last term in eq.~\ref{eq:hamiltonian}, $V_{XC}$, is the exchange-correlation potential, which is approximated by the adiabatic local-density approximation (ALDA), obtained from a parameterization to a homogeneous electron gas by Perdew and Zunger~\cite{PhysRevB.23.5048}. 

At the beginning of the TDDFT calculations, the ground state of the system is prepared by performing a density-functional theory (DFT) calculation. With these initial conditions in place, we then proceed to propagate the KS orbitals, $\psi_{k}(\mathbf{r},t)$ over time by using the time-dependent KS equation, given as 
\begin{equation}
i \frac{\partial \psi_k(\mathbf{r}, t)}{\partial t} = \hat{H}_{\text{KS}}(t) \psi_k(\mathbf{r}, t).
\label{eq:tdks}
\end{equation}
Eq.~\ref{eq:tdks} was solved using the following time propagator
\begin{equation}
\psi_k(\mathbf{r}, t + \delta t) = \exp\left(\frac{-i \delta t}{\hbar} \hat{H}_{\text{KS}}(t) \right) \psi_k(\mathbf{r}, t).
\end{equation}
This operator is approximated using a fourth-degree Taylor expansion, given as
\begin{equation}
\psi_k(\mathbf{r}, t + \delta t) \approx \sum_{n=0}^{4} \frac{1}{n!} \left(\frac{-i \delta t}{\hbar} \hat{H}_{\text{KS}}(t)\right)^n \psi_k(\mathbf{r}, t).
\end{equation}
The operator is applied for $N$ time steps until the final time, $t_{final} = N \cdot \delta t$, is obtained. 
The Taylor propagation is conditionally stable, and the time step has
to satisfy \cite{Varga_Driscoll_2011a} $\delta t<1.87(\Delta x/\pi)^2$.
For a typical grid spacing of $\Delta x = 0.3$ \AA\  this means that the
timestep must be smaller than 1.4 as. In our simulations, a time step of $\delta t = 1$~attosecond (as) and a 
total propagation time of $t_{\text{final}} = 120$~femtoseconds (fs) were used.

In real-space TDDFT, the KS orbitals are represented at discrete points on a uniform 
rectangular grid in real space. The simulation accuracy is governed by the grid spacing. In our calculations, 
we employed a grid spacing 
of 0.3 \AA\ and used 100 grid points along each of the $x$-, $y$-, and $z$-axes.

To enforce boundary conditions, we set the KS orbitals to zero at the edges of the simulation cell. However, during proton impact events, the collision can impart sufficient energy to the electronic wavefunctions, potentially leading to ionization or the ejection of electronic density beyond the molecule. In such cases, unphysical reflections of the wavefunction from the cell boundaries can occur, introducing artifacts into the simulation. To mitigate this issue, we implemented a complex absorbing potential (CAP) to dampen the wavefunction as it approaches the boundaries. The specific form of the CAP used in our simulations, as described by Manolopoulos~\cite{10.1063/1.1517042}, is given by:
\begin{equation}
- i w(x) = -i \frac{\hbar^2}{2m} \left(\frac{2\pi}{\Delta x}\right)^2 f(y),
\end{equation}
where $x_{1}$ is the start and $x_{2}$ is the end of the absorbing region, $\Delta x = x_{2} - x_{1}$, $c = 2.62$ is a numerical constant, $m$ is the electron’s mass, and
\begin{equation}
f(y) = \frac{4}{c^2} \left( \frac{1}{(1 + y)^2} + \frac{1}{(1 - y)^2} - 2 \right), \quad y = \frac{x - x_1}{\Delta x}.
\end{equation}

As the molecule becomes ionized during the simulation, electron density is driven towards the CAP. Additionally, any ejected fragments carry their associated electron density as they move towards the boundaries. When electron density reaches the CAP region, it is absorbed. Consequently, the total number of electrons,
\begin{equation}
N(t) = \int_V \rho(\mathbf{r}, t) \, d^3x,
\end{equation}
where \(V\) is the volume of the simulation box, decreases relative to the initial electron number, \(N(0)\). We interpret \(N(0) - N(t)\) as the total number of electrons that have been ejected from the simulation box.

Motion of the ions in the simulations were treated classically. Using the Ehrenfest theorem, 
the quantum forces on the ions due to the electrons are given by the derivatives 
of the expectation value of the total electronic energy with respect to the ionic positions. 
These forces are then fed into Newton’s Second Law, giving
\begin{equation}
\begin{split}
M_i \frac{d^2 \mathbf{R}_i}{dt^2} = \sum_{j \neq i}^{N_{\text{ions}}} \frac{Z_i Z_j (\mathbf{R}_i - \mathbf{R}_j)}{|\mathbf{R}_i - \mathbf{R}_j|^3}\\ 
- \nabla_{\mathbf{R}_i} \int V_{\text{ion}}(\mathbf{r}, \mathbf{R}_i) \rho(\mathbf{r}, t) \, d\mathbf{r},
\end{split}
\end{equation}
where $M_{i}$, $Z_{i}$, and $\mathbf{R}_{i}$ are the mass, 
pseudocharge (valence), and position of the $i^{th}$ ion, respectively, and $N_{\text{ions}}$ is the total number of ions. 
$V_{\text{ion}}(\mathbf{r},\mathbf{R}_i)$ is the pseudopotential representing the combined effect of the nucleus and core electrons, and it interacts with the electron density $\rho(\mathbf{r}, t)$ via Ehrenfest dynamics.
This differential equation was time propagated using the Verlet algorithm at every time step $\delta t$.

\begin{figure}[ht!]
    \centering
    \includegraphics[width=0.95\columnwidth]{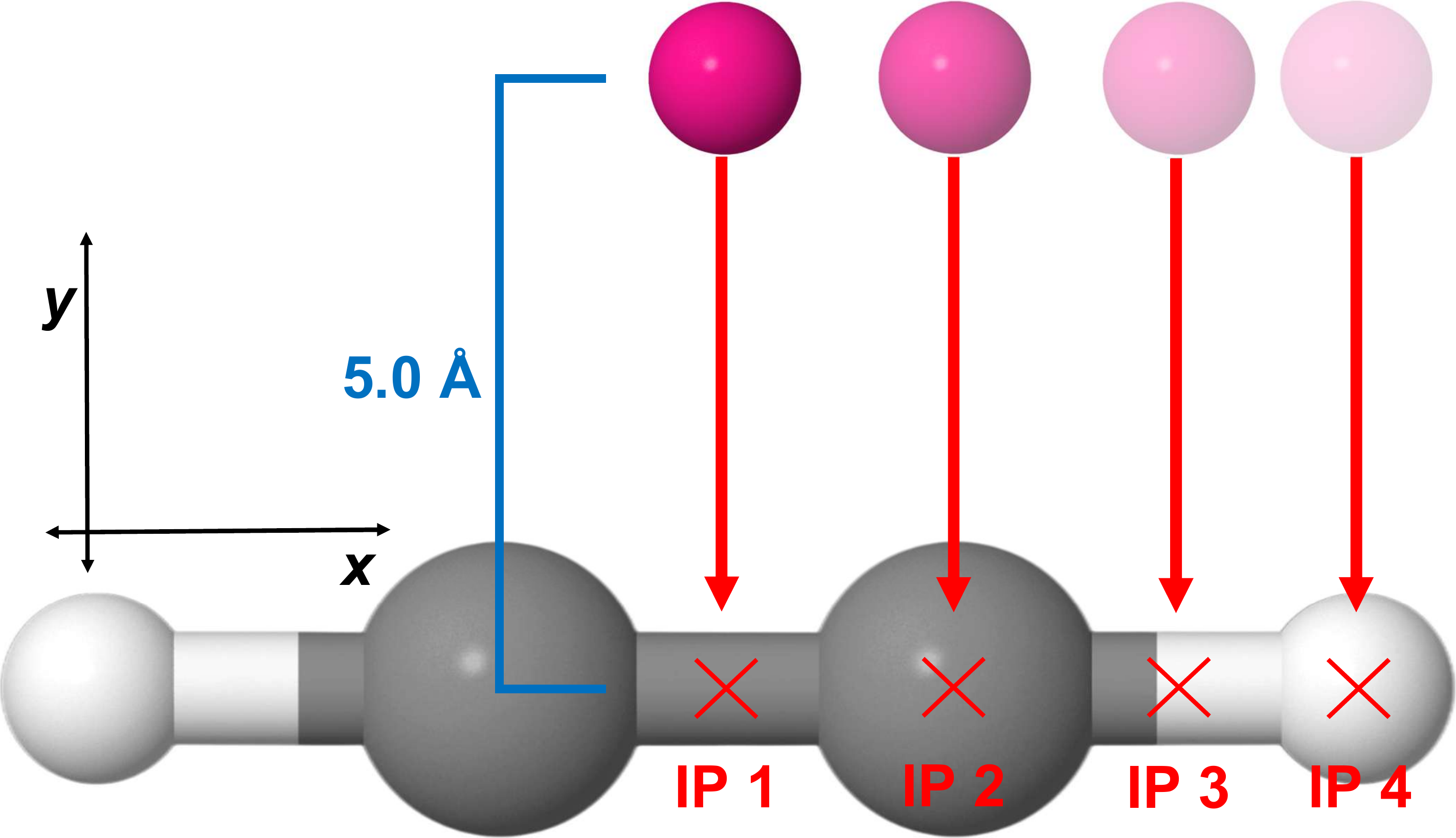}
    \caption{Impact point (IP) diagram for C\textsubscript{2}H\textsubscript{2}. The initial proton positions (shown in pink) corresponding to different IPs are placed 5.0~\AA\ above the molecular plane. Distances are not drawn to scale.}
    \label{fig:c2h2-ip-diagram}
\end{figure}

The target hydrocarbon molecule was placed such that the center of its equilibrium geometry 
coincided with the origin, with its longest molecular axis aligned along the $x$-axis. 
The incident proton was initially positioned 5.0~\AA\ above the target, orthogonal 
to the molecule’s largest cross-sectional area 
(see Fig.~\ref{fig:c2h2-ip-diagram}). At the chosen separation
distance, the interaction between the proton and molecule is
negligible. Positioning the proton farther away would not alter
the physical results but would require a larger simulation cell,
leading to significantly higher computational demands and longer
calculation times.
Both the target and the proton were contained within a 
29.7~\AA\ $\times$ 29.7~\AA\ $\times$ 29.7~\AA\ simulation grid centered at 
the origin, providing sufficient space along the proton’s trajectory to 
capture the full dynamics of the collision. The simulations were performed at 
zero temperature to remove thermal motion of the target atoms as a variable. The atomic nuclei 
were allowed to move from the forces experienced during the collision, since nuclear 
motion and energy redistribution into vibrational modes strongly influence fragmentation, 
bonding, and scattering outcomes. 

It should be emphasized that these simulations represent an idealized limit. In experimental conditions, target molecules possess finite temperature, undergo random motion, and can be struck at a wide distribution of incident angles and positions. Capturing such effects computationally would require an extensive sampling over molecular orientations, impact geometries, and thermal ensembles, greatly increasing the computational cost. Our approach therefore represents a balance: by focusing on equilibrium geometries, impact points of interest, and orthogonal incidence, we isolate and probe the most characteristic features of proton–hydrocarbon collisions, while recognizing that real experiments will exhibit additional statistical broadening and variability.

In the following section, we present the results of proton collisions with three hydrocarbons of varying size: acetylene (C\textsubscript{2}H\textsubscript{2}), propane (C\textsubscript{3}H\textsubscript{8}), and butane (C\textsubscript{4}H\textsubscript{10}). For each molecule, impact points were chosen at chemically relevant sites, including C--C bonds, C atoms, C-H bonds, and H atoms. This selection provides a representative sampling of key collision sites while keeping the computational cost manageable. At each impact point, seven proton initial kinetic energies were considered: 0.52, 4.66, 12.96, 25.39, 41.98, 62.70, and 87.58~eV, motivated by the typical kinetic energies of protons produced 
in Coulomb explosion experiments~\cite{PhysRevLett.106.163001,PhysRevLett.92.063001}.

\section{Results}

\subsection{Acetylene (C\textsubscript{2}H\textsubscript{2})}
\label{subsec:acetylene}

The selected impact points (IPs) for proton collisions with C\textsubscript{2}H\textsubscript{2} are illustrated in Fig.~\ref{fig:c2h2-ip-diagram}. Due to molecular symmetry, only incident points located at or to the right of the molecular center were considered. Specifically, IP~1, IP~2, IP~3, and IP~4 correspond to the central C--C bond, the right C atom, the right C–H bond, and the terminal H atom of C\textsubscript{2}H\textsubscript{2}, respectively.

\begin{table*}[t]
\centering
\renewcommand{\arraystretch}{1.2}
\setcellgapes{2pt}\makegapedcells
\begin{tabular}{lccccccc}
\toprule
\makecell{\textbf{Impact Point (IP)}\\\textit{Proton Kinetic Energy (eV)}} & \textbf{0.52} & \textbf{4.66} & \textbf{12.96} & \textbf{25.39} & \textbf{41.98} & \textbf{62.70} & \textbf{87.58} \\
\midrule
\arrayrulecolor[gray]{0.75}
\textbf{IP 1} & \makecell{S, 0.26 \\ $\mathrm{C_2H_2}^{1.01+}$ \\ $\mathrm{H}^{0.25+}$} & 
\makecell{S, 3.13 \\ $\mathrm{C_2H_2}^{0.65+}$ \\ $\mathrm{H}^{0.63+}$} &
\makecell{S, 7.35 \\ $\mathrm{C_2H_2}^{0.63+}$ \\ $\mathrm{H}^{0.71+}$} &
\makecell{S, 13.47 \\ $\mathrm{C_2H_2}^{0.79+}$ \\ $\mathrm{H}^{0.54+}$} &
\makecell{S, 12.66 \\ $\mathrm{C_2H_2}^{0.87+}$ \\ $\mathrm{H}^{0.44+}$} &
\makecell{S, 8.73 \\ $\mathrm{C_2H_2}^{0.81+}$ \\ $\mathrm{H}^{0.56+}$} &
\makecell{S, 8.82 \\ $\mathrm{C_2H_2}^{0.90+}$ \\ $\mathrm{H}^{0.44+}$} \\
\midrule
\textbf{IP 2} & \makecell{P, -- \\ $\mathrm{C_2H_3}^{1.24+}$} &
\makecell{S, 2.21 \\ $\mathrm{C_2H_2}^{0.74+}$ \\ $\mathrm{H}^{0.50+}$} &
\makecell{S, 4.39 \\ $\mathrm{C_2H_2}^{0.78+}$ \\ $\mathrm{H}^{0.42+}$} &
\makecell{S, 8.08 \\ $\mathrm{C_2H_2}^{0.80+}$ \\ $\mathrm{H}^{0.40+}$} &
\makecell{S, 13.00 \\ $\mathrm{C_2H_2}^{0.51+}$ \\ $\mathrm{H}^{0.67+}$} &
\makecell{S, 19.90 \\ $\mathrm{C_2H_2}^{0.65+}$ \\ $\mathrm{H}^{0.58+}$} &
\makecell{S, 29.22 \\ $\mathrm{C_2H_2}^{0.79+}$ \\ $\mathrm{H}^{0.53+}$} \\
\midrule
\textbf{IP 3} & \makecell{P, -- \\ $\mathrm{C_2H_3}^{1.22+}$} &
\makecell{S, 3.30 \\ $\mathrm{C_2H_2}^{0.66+}$ \\ $\mathrm{H}^{0.56+}$} &
\makecell{S, 10.56 \\ $\mathrm{C_2H}^{0.30+}$ \\ $\mathrm{H}^{0.44+}$ \\ $\mathrm{H}^{0.43+}$} &
\makecell{A, -- \\ $\mathrm{C_2H_2}^{0.75+}$ \\ $\mathrm{H}^{0.50+}$} &
\makecell{S, 12.76 \\ $\mathrm{C_2H}^{0.30+}$ \\ $\mathrm{H}^{0.39+}$ \\ $\mathrm{H}^{0.42+}$} &
\makecell{S, 7.87 \\ $\mathrm{C_2H_2}^{0.74+}$ \\ $\mathrm{H}^{0.49+}$} &
\makecell{S, 6.81 \\ $\mathrm{C_2H_2}^{0.82+}$ \\ $\mathrm{H}^{0.41+}$} \\
\midrule
\textbf{IP 4} & \makecell{P, -- \\ $\mathrm{C_2H_3}^{1.20+}$} &
\makecell{A, -- \\ $\mathrm{C_2H_2}^{0.61+}$ \\ $\mathrm{H}^{0.54+}$} &
\makecell{A, -- \\ $\mathrm{C_2H_2}^{0.49+}$ \\ $\mathrm{H}^{0.58+}$} &
\makecell{A, -- \\ $\mathrm{C_2H_2}^{0.80+}$ \\ $\mathrm{H}^{0.30+}$} &
\makecell{A, -- \\ $\mathrm{C_2H_2}^{0.55+}$ \\ $\mathrm{H}^{0.58+}$} &
\makecell{A, -- \\ $\mathrm{C_2H_2}^{0.57+}$ \\ $\mathrm{H}^{0.55+}$} &
\makecell{A, -- \\ $\mathrm{C_2H_2}^{0.73+}$ \\ $\mathrm{H}^{0.38+}$} \\
\arrayrulecolor{black}
\bottomrule
\end{tabular}
\caption{Combined outcome data for different C\textsubscript{2}H\textsubscript{2} IPs (rows) and proton KEs (columns). Each cell indicates the reaction type (S: scattering, P: proton capture, A: abstraction), the KE loss of the proton, and the resulting fragment products with their respective charges.}
\label{tab:c2h2-data}
\end{table*}

The outcomes of proton impacts on C\textsubscript{2}H\textsubscript{2} are summarized in Table~\ref{tab:c2h2-data}. Each row of the table corresponds to one of the four selected IPs (IP~1–4, labeled in the first column), while the subsequent columns present the results for different initial kinetic energies (KEs) of the proton projectile, ranging from 0.52 to 87.58~eV.

In Table~\ref{tab:c2h2-data}, each cell represents the outcome of a specific simulation defined by a chosen IP and an initial proton KE. The first entry in each cell denotes the reaction type. Three distinct outcomes were observed under the present conditions: ``S'' (scattering), ``P'' (proton capture), and ``A'' (abstraction). ``S'' indicates cases where the proton was unable to form a bond and was instead reflected or scattered. Scattering events may result in molecular fragmentation as well. ``P'' corresponds to proton capture, in which the proton was retained by the molecule and formed a stable bond throughout the simulation window with no fragmentation occurring. ``A'' denotes abstraction events, where the proton induced molecular fragmentation or the dissociation of a single atom, and subsequently bonded with the molecule or one of its fragments. Following the reaction type, the proton’s KE loss is reported. This value is only present for scattering events, since the proton retains some KE as it departs from the molecule. In the case of proton capture or abstraction, this entry is marked by ``--'', as the proton becomes bonded and effectively loses its KE. The subsequent lines within each cell list the final molecular states and fragments, specifying the reaction products and their corresponding charge states.

Analyzing each cell in Table~\ref{tab:c2h2-data} provides 
information about the outcome of a given IP and proton KE. For example, consider 
the case of IP~2 with an initial proton KE of 41.98~eV. The reaction is classified as scattering (denoted by ``S''), with the proton losing 13.00~eV of its KE. In addition, charge analysis shows that the proton captured 0.33 electrons from C\textsubscript{2}H\textsubscript{2}, as reflected in its final charge state of H$^{0.67+}$ (reduced from the initial H$^{1.00+}$). Meanwhile, the hydrocarbon target (C\textsubscript{2}H\textsubscript{2}) exhibits a final charge state of 0.51+, corresponding to a net loss of 0.51 electrons. This indicates that, during the scattering event, electron ejection further ionized C\textsubscript{2}H\textsubscript{2}: 
0.33 electrons were transferred from the molecule to the proton, 
while an additional 0.18 electrons were emitted into the CAP  
(the meaning of fractional charges observed in molecular fragments will
be explained in subsequent discussion). Thus, even in scattering events, rich dynamics emerge involving 
simultaneous electron capture, ionization, and the transfer of KE from the 
projectile into both nuclear and electronic degrees of freedom of the target molecule.

\begin{figure*}
\centering
\includegraphics[width=\textwidth]{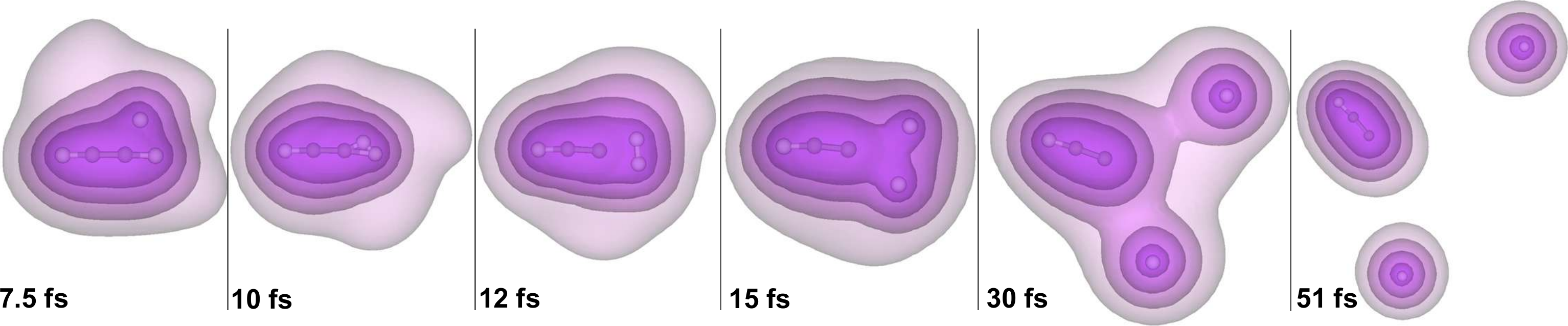}
\caption{Scattering dynamics of a proton incident on C\textsubscript{2}H\textsubscript{2}, directed toward IP~3 with an initial KE of 12.96~eV. The electron density isosurfaces are shown in purple at values of 0.5, 0.1, 0.001, and 0.0001.}
\label{fig:c2h2-ip3-v-0.5}
\end{figure*}

In certain cases, scattering events are accompanied by molecular fragmentation. For example, in Table~\ref{tab:c2h2-data}, the simulation corresponding to IP~3 with an initial proton KE of 12.96~eV produces the fragments C\textsubscript{2}H$^{0.30+}$, H$^{0.44+}$, and H$^{0.43+}$. In all scattering cases that yield multiple fragments, the final entry corresponds to the proton projectile (here, H$^{0.43+}$), while the preceding H ion (H$^{0.44+}$) originates from the target molecule. In this instance, the projectile proton dislodged an H atom from the C\textsubscript{2}H\textsubscript{2}, ionizing it in the process and simultaneously capturing electrons, which led to the charge states listed in the table. The proton lost 10.56~eV of KE during this interaction. Snapshots of the molecular trajectories and electron densities for this event are shown in Fig.~\ref{fig:c2h2-ip3-v-0.5}. Between 7.5 and 10~fs, the proton approaches IP~3 (the right C–H bond) and wedges itself between the C and H atoms. The resulting close approach generates strong Coulomb repulsion, expelling the terminal H atom downward while deflecting the proton rightward (10–12~fs). Subsequently, mutual repulsion between the positively charged proton and ejected H ion further increases their separation (12–51~fs). By 51~fs, the system consists of a CH\textsubscript{2} fragment and two separated H ions, consistent with the products identified in the table.

Analyzing a case that results in proton capture is equally insightful. For instance, at IP~2 with an initial proton KE of 0.52~eV (Table~\ref{tab:c2h2-data}), the proton is captured by the molecule, forming a stable bond as reflected in the final molecular state C\textsubscript{2}H\textsubscript{3}$^{1.24+}$. The final charge state also indicates electron ejection. Since the C\textsubscript{2}H\textsubscript{2} molecule begins neutral and the incident proton carries a charge of 1+, the expected charge of the protonated product would be 1+ if no electrons were lost. The observed charge of 1.24+ instead implies that an additional 0.24 electrons were emitted from the hydrocarbon, revealing that electron ejection accompanied the capture process.

\begin{figure*}
\centering
\includegraphics[width=\textwidth]{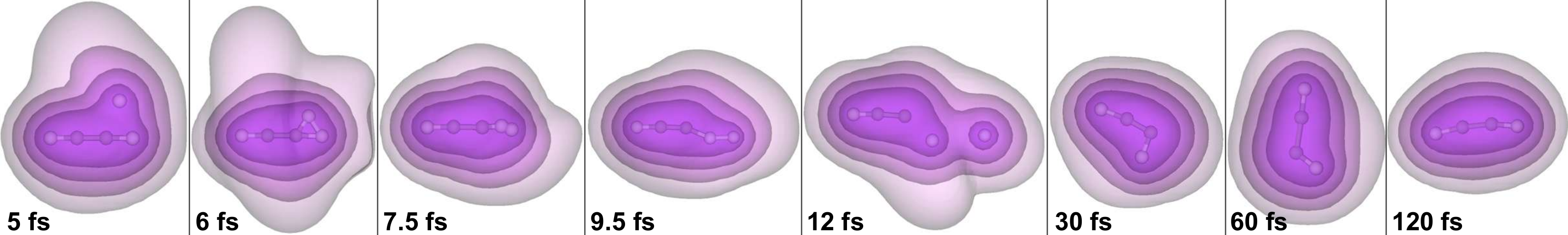}
\caption{Abstraction event dynamics of a proton incident on C\textsubscript{2}H\textsubscript{2}, directed toward IP~3 with an initial KE of 25.39~eV. The electron density isosurfaces are shown in purple at values of 0.5, 0.1, 0.001, and 0.0001.}
\label{fig:c2h2-ip3-v-0.7}
\end{figure*}

The final reaction pathway is illustrated for IP~3 at an initial proton KE of 25.39~eV, as shown in Fig.~\ref{fig:c2h2-ip3-v-0.7}. In this trajectory, the proton approaches the C\textsubscript{2}H\textsubscript{2} molecule between 5 and 7.5~fs, inserting itself between a C and H atom. Between 7.5 and 9.5~fs, the molecular framework elongates to accommodate the incoming proton, accompanied by strong interatomic repulsion. By 12~fs, the projectile transfers its KE, displacing the terminal H atom and ejecting it from the molecule, while remaining bound near the C\textsubscript{2}H fragment. From 30 to 120~fs, the proton establishes a stable bond within the hydrocarbon, yielding a C\textsubscript{2}H\textsubscript{2} fragment. Throughout this process, the molecule exhibits pronounced rotational motion, indicating that a portion of the proton’s initial KE is converted into rotation, which contributes to stabilizing the newly formed bond.

As shown in Table~\ref{tab:c2h2-data}, the case analyzed in Fig.~\ref{fig:c2h2-ip3-v-0.7} is particularly notable because it represents the only instance at IP~3 that leads to abstraction. Adjusting the initial KE either higher or lower instead results in scattering. This behavior highlights the existence of a critical KE range necessary to inject the proton into the molecular system without it being scattered. At this energy, the KE is not so high that the proton simply passes through the C-H bond without being captured, nor is it so low that the proton approaches too close to the C and H nuclear cores and is repelled, as illustrated in Fig.~\ref{fig:c2h2-ip3-v-0.5}. Instead, at approximately 25.39~eV, the proton can effectively bind to the hydrocarbon, displacing an H atom in the process.

Interestingly, abstraction is particularly prevalent at IP~4, occurring in every calculation with initial KEs greater than or equal to 4.66~eV. In the case of 4.66~eV, the proton first fragments the molecule by detaching one of the hydrogens\textemdash observed as the H$^{0.54+}$ fragment\textemdash and subsequently bonds with the remaining C\textsubscript{2}H fragment, forming the stable C\textsubscript{2}H\textsubscript{2}$^{0.61+}$ listed in Table~\ref{tab:c2h2-data}. This process effectively replaces the H atom originally bound to the hydrocarbon. The total charge of the resulting fragments is 1.15+, indicating that approximately 0.15 electrons were ejected from the system during the abstraction event.
 
While individual calculations across various IPs and proton KEs provide valuable insights, broader trends emerge when comparing results as a function of IP and proton KE in Table~\ref{tab:c2h2-data}. For example, at IP~1 the outcome is always scattering, regardless of the proton KE. This is reasonable given that IP~1 lies at the center of the C--C bond (see Fig.~\ref{fig:c2h2-ip-diagram}). In hydrocarbons, hydrogen preferentially bonds in linear or pyramidal geometries at positions maximally separated from neighboring atoms. Here, however, the proton is introduced too close to both carbon nuclei, leaving insufficient space to form a bond and resulting in rapid ejection due to the strong Coulombic repulsion from the carbon cores.  

In contrast, at IP~4 the dominant process is abstraction whenever the incoming proton has an initial KE of or above 4.66~eV. In these cases, the proton replaces the right H atom in the C\textsubscript{2}H\textsubscript{2} structure (see Fig.~\ref{fig:c2h2-ip-diagram}). This high probability is consistent with geometry: IP~4 is located directly at the position of the right H atom. The proton arrives with sufficient KE to overcome the local potential barrier and approach the target nucleus. As it nears the H atom, Coulombic repulsion with the hydrocarbon nuclei decelerates the proton, transferring its KE to the target H atom and breaking its bond. The displaced H atom is ejected, while the proton slows enough to stabilize and occupy its position, thereby completing the abstraction process.  

This mechanism is consistent across all tested KEs of and above 4.66~eV at IP~4. Even at very high values, such as 87.58~eV, nearly all of the proton’s KE is transferred to the target H atom, which is knocked out, while the proton itself comes to rest and bonds to the hydrocarbon. At the lowest tested KE (0.52~eV), the proton approaches slowly enough to allow the target H atom to shift and make space for the proton to bond. The relatively small KE can then be redistributed among nuclear and electronic degrees of freedom, consistent with the 0.20 electron ejection reported in Table~\ref{tab:c2h2-data}.

In Table~\ref{tab:c2h2-data}, the reaction outcome is seen to depend
primarily on the IP, as most rows display consistent behavior across
all KEs, with only one or two exceptions per row. Nevertheless, KE
still plays an important role. At sufficiently low KE (0.52 eV), when the proton does not strike the central C--C bond (IP 1), proton capture occurs at the remaining incident points (IP 2–4). In this regime, the low KE enables local atomic rearrangements that facilitate bond formation while limiting energy redistribution across the molecular degrees of freedom, including charge redistribution.

\subsection{Propane (C\textsubscript{3}H\textsubscript{8})}
\label{subsec:propane}

\begin{figure}[ht!]
    \centering
    \includegraphics[width=0.95\columnwidth]{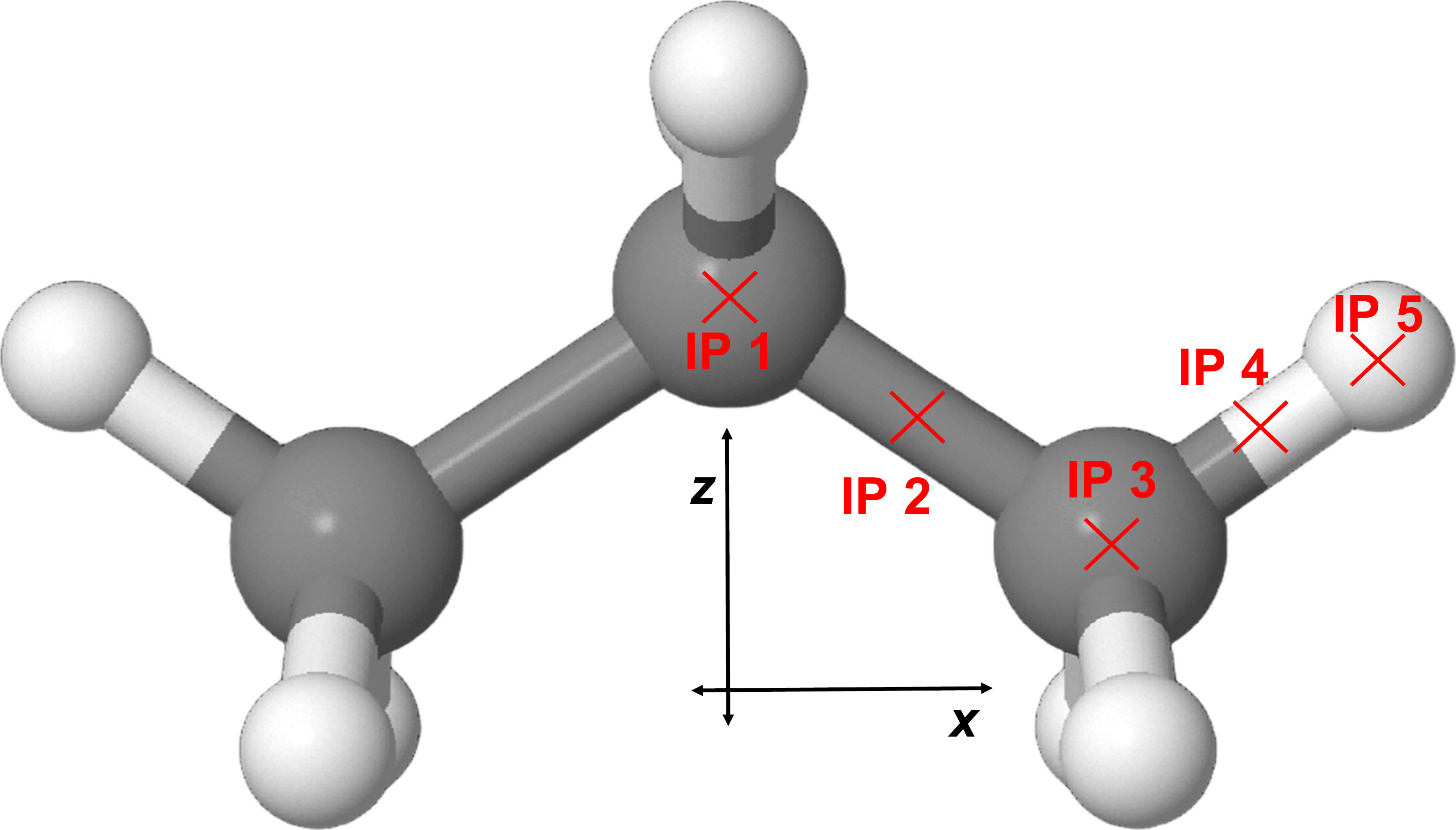}
    \caption{Impact point (IP) diagram for C\textsubscript{3}H\textsubscript{8}.}
    \label{fig:c3h8-ip-diagram}
\end{figure}

The selected IPs for proton collisions with C\textsubscript{3}H\textsubscript{8} are illustrated in Fig.~\ref{fig:c3h8-ip-diagram}. Due to molecular symmetry, only incident points at and to the right of the molecular center are considered. IP~1 through IP~5 correspond, respectively, to the central C atom, the right C--C bond, the right C atom, the right C–H bond, and the terminal H atom of the C\textsubscript{3}H\textsubscript{8} molecule.

\begin{table*}[t]
\centering
\renewcommand{\arraystretch}{1.2}
\setcellgapes{2pt}\makegapedcells
\begin{tabular}{lccccccc}
\toprule
\makecell{\textbf{Impact Point (IP)}\\\textit{Proton Kinetic Energy (eV)}} & \textbf{0.52} & \textbf{4.66} & \textbf{12.96} & \textbf{25.39} & \textbf{41.98} & \textbf{62.70} & \textbf{87.58} \\
\midrule
\arrayrulecolor[gray]{0.75}
\textbf{IP 1} & \makecell{S, -0.14 \\ $\mathrm{C_3H_7}^{1.14+}$ \\ $\mathrm{H}^{0.01+}$ \\ $\mathrm{H}^{0.07+}$} & \makecell{S, 3.09 \\ $\mathrm{C_3H_7}^{1.12+}$ \\ $\mathrm{H}^{0.10+}$ \\ $\mathrm{H}^{0.07+}$} & \makecell{S, 6.29 \\ $\mathrm{C_3H_8}^{1.01+}$ \\ $\mathrm{H}^{0.35+}$} & \makecell{S, 11.05 \\ $\mathrm{C_3H_8}^{1.05+}$ \\ $\mathrm{H}^{0.33+}$} & \makecell{S, 16.42 \\ $\mathrm{C_3H_8}^{1.12+}$ \\ $\mathrm{H}^{0.30+}$} & \makecell{S, 23.27 \\ $\mathrm{C_3H_7}^{1.12+}$ \\ $\mathrm{H}^{0.03-}$ \\ $\mathrm{H}^{0.32+}$} & \makecell{S, 30.73 \\ $\mathrm{CH_2}^{0.29+}$ \\ $\mathrm{CH_3}^{0.31+}$ \\ $\mathrm{CH_3}^{0.47+}$ \\ $\mathrm{H}^{0.32+}$} \\
\midrule
\textbf{IP 2} & \makecell{A, -- \\ $\mathrm{C_3H_7}^{1.17+}$ \\ $\mathrm{H_2}^{0.05+}$} & \makecell{S, 4.54 \\ $\mathrm{C_3H_7}^{1.24+}$ \\ $\mathrm{H}^{0.07+}$ \\ $\mathrm{H}^{0.04+}$} & \makecell{S, 12.59 \\ $\mathrm{C_2H_5}^{0.70+}$ \\ $\mathrm{CH_3}^{0.40+}$ \\ $\mathrm{H}^{0.23+}$} & \makecell{S, 7.93 \\ $\mathrm{C_3H_8}^{0.92+}$ \\ $\mathrm{H}^{0.47+}$} & \makecell{S, 7.04 \\ $\mathrm{C_3H_8}^{0.96+}$ \\ $\mathrm{H}^{0.42+}$} & \makecell{S, 6.90 \\ $\mathrm{C_3H_8}^{0.91+}$ \\ $\mathrm{H}^{0.42+}$} & \makecell{S, 6.22 \\ $\mathrm{C_3H_8}^{0.88+}$ \\ $\mathrm{H}^{0.46+}$} \\
\midrule
\textbf{IP 3} & \makecell{P, -- \\ $\mathrm{C_3H_9}^{1.25+}$} & \makecell{S, 3.72 \\ $\mathrm{C_3H_7}^{1.23+}$ \\ $\mathrm{H}^{0.11+}$ \\ $\mathrm{H}^{0.05-}$} & \makecell{S, 6.58 \\ $\mathrm{C_3H_8}^{0.96+}$ \\ $\mathrm{H}^{0.39+}$} & \makecell{S, 10.16 \\ $\mathrm{C_3H_8}^{0.95+}$ \\ $\mathrm{H}^{0.38+}$} & \makecell{S, 16.15 \\ $\mathrm{C_2H_5}^{0.67+}$ \\ $\mathrm{CH_3}^{0.35+}$ \\ $\mathrm{H}^{0.36+}$} & \makecell{S, 23.13 \\ $\mathrm{C_2H_5}^{0.75+}$ \\ $\mathrm{CH_3}^{0.30+}$ \\ $\mathrm{H}^{0.34+}$} & \makecell{S, 30.34 \\ $\mathrm{C_2H_5}^{0.68+}$ \\ $\mathrm{CH_2}^{0.43+}$ \\ $\mathrm{H}^{0.09-}$ \\ $\mathrm{H}^{0.34+}$} \\
\midrule
\textbf{IP 4} & \makecell{P, -- \\ $\mathrm{C_3H_9}^{1.25+}$} & \makecell{S, 4.43 \\ $\mathrm{C_3H_7}^{1.16+}$ \\ $\mathrm{H}^{0.03+}$ \\ $\mathrm{H}^{0.11+}$} & \makecell{A, -- \\ $\mathrm{C_3H_7}^{1.07+}$ \\ $\mathrm{H_2}^{0.22+}$} & \makecell{S, 17.22 \\ $\mathrm{C_3H_7}^{1.06+}$ \\ $\mathrm{H}^{0.15+}$ \\ $\mathrm{H}^{0.13+}$} & \makecell{S, 12.68 \\ $\mathrm{C_3H_7}^{1.00+}$ \\ $\mathrm{H}^{0.12+}$ \\ $\mathrm{H}^{0.23+}$} & \makecell{S, 10.16 \\ $\mathrm{C_3H_7}^{0.82+}$ \\ $\mathrm{H}^{0.19+}$ \\ $\mathrm{H}^{0.33+}$} & \makecell{S, 8.69 \\ $\mathrm{C_3H_7}^{1.06+}$ \\ $\mathrm{H}^{0.03-}$ \\ $\mathrm{H}^{0.33+}$} \\
\midrule
\textbf{IP 5} & \makecell{P, -- \\ $\mathrm{C_3H_9}^{1.22+}$} & \makecell{A, -- \\ $\mathrm{C_3H_8}^{0.79+}$ \\ $\mathrm{H}^{0.45+}$} & \makecell{A, -- \\ $\mathrm{C_3H_8}^{0.86+}$ \\ $\mathrm{H}^{0.36+}$} & \makecell{A, -- \\ $\mathrm{C_3H_8}^{0.86+}$ \\ $\mathrm{H}^{0.37+}$} & \makecell{A, -- \\ $\mathrm{C_3H_8}^{0.88+}$ \\ $\mathrm{H}^{0.39+}$} & \makecell{A, -- \\ $\mathrm{C_3H_8}^{0.88+}$ \\ $\mathrm{H}^{0.38+}$} & \makecell{A, -- \\ $\mathrm{C_3H_8}^{0.72+}$ \\ $\mathrm{H}^{0.46+}$} \\
\arrayrulecolor{black}
\bottomrule
\end{tabular}
\caption{Combined outcome data for different C\textsubscript{3}H\textsubscript{8} IPs (rows) and proton KEs (columns). Each cell shows the reaction type (S: scattering, P: proton capture, A: abstraction), followed by the KE loss of the proton and the resulting fragment products with their respective charges.}
\label{tab:c3h8-data}
\end{table*}

The results for C\textsubscript{3}H\textsubscript{8}, including the reaction type, proton KE loss, and fragment products across the various KEs and IPs, are summarized in Table~\ref{tab:c3h8-data}. The format and types of information presented\textemdash namely scattering, proton capture, and abstraction\textemdash are consistent with those analyzed for C\textsubscript{2}H\textsubscript{2} in Sec.~\ref{subsec:acetylene} and Table~\ref{tab:c2h2-data}.

\begin{figure*}
\centering
\includegraphics[width=\textwidth]{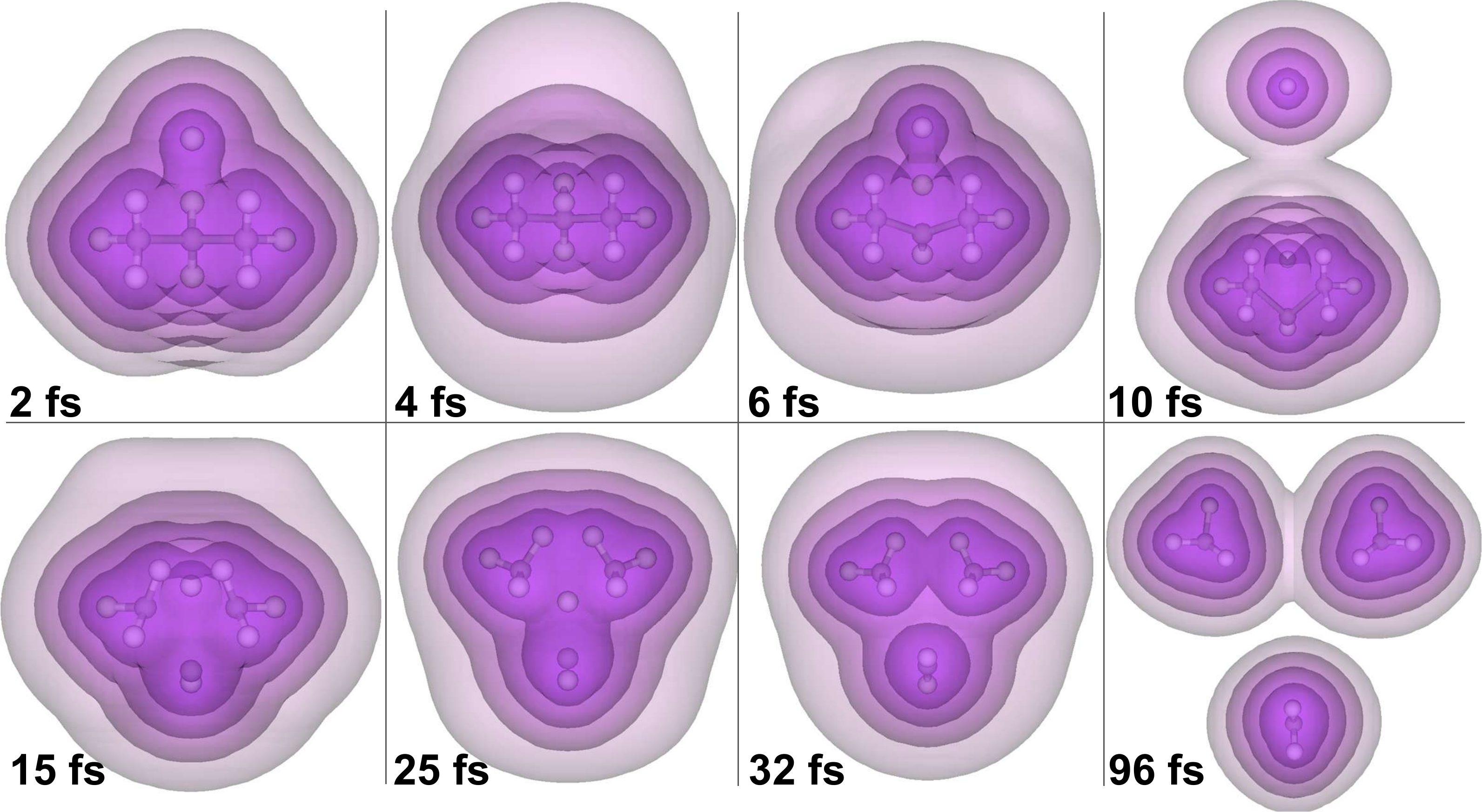}
\caption{Scattering dynamics of a proton incident on C\textsubscript{3}H\textsubscript{8}, directed toward IP~1 with an initial KE of 87.58~eV. The electron density isosurfaces are shown in purple at values of 0.5, 0.1, 0.001, and 0.0001.}
\label{fig:c3h8-ip1-v-1.3}
\end{figure*}

As shown in Table~\ref{tab:c3h8-data}, the collision dynamics for C\textsubscript{3}H\textsubscript{8} can produce particularly striking outcomes. Due to the molecule's larger size, proton-induced fragmentation can be extensive. For example, at IP~1 with an initial KE of 87.58~eV, the resulting fragments are CH\textsubscript{2}$^{0.29+}$, CH\textsubscript{3}$^{0.31+}$, CH\textsubscript{3}$^{0.47+}$, and H$^{+0.32}$ (the proton projectile). Snapshots of the molecular breakup is illustrated in Fig.~\ref{fig:c3h8-ip1-v-1.3}. The proton approaches the molecule at 2~fs and reaches its closest distance at 4~fs, after which it is rapidly repelled and scattered away by 6~fs. This energy transfer causes the C\textsubscript{3}H\textsubscript{8} molecular geometry to bend and recoil, as indicated by the central C atom dipping between 6~fs and 10~fs. By 15~fs, the proton has fully separated from the molecule, leaving two CH\textsubscript{3} fragments on either side, a C–H bond at the bottom, and an unbound hydrogen in the center.  

Subsequently, asymmetric charge distribution and Coulomb repulsion among the positively charged CH\textsubscript{3} fragments drive hydrogen migration toward the CH fragment at the bottom, forming a CH\textsubscript{2} fragment by 32~fs. The atoms remain bound but continue to repel one another due to their positive charge, as observed at 96~fs. This trajectory demonstrates how a single proton collision can produce multiple fragments, including hydrogen migration events. In the context of Coulomb explosion, the occurrence of secondary fragmentation (summarized in Table~\ref{tab:c3h8-data}) shows that some molecular products can result solely from proton–molecule collisions rather than from direct laser-induced ionization. Such products may also appear among the fragment distributions reported in Coulomb explosion experiments~\cite{PhysRevA.111.013109,Last2002,Cornaggia_1992,Mogyorosi2025,PhysRevLett.92.063001,D3CP02337K,PhysRevA.111.033109,PhysRevA.82.043433,Livshits2006}.

\begin{figure*}
\centering
\includegraphics[width=\textwidth]{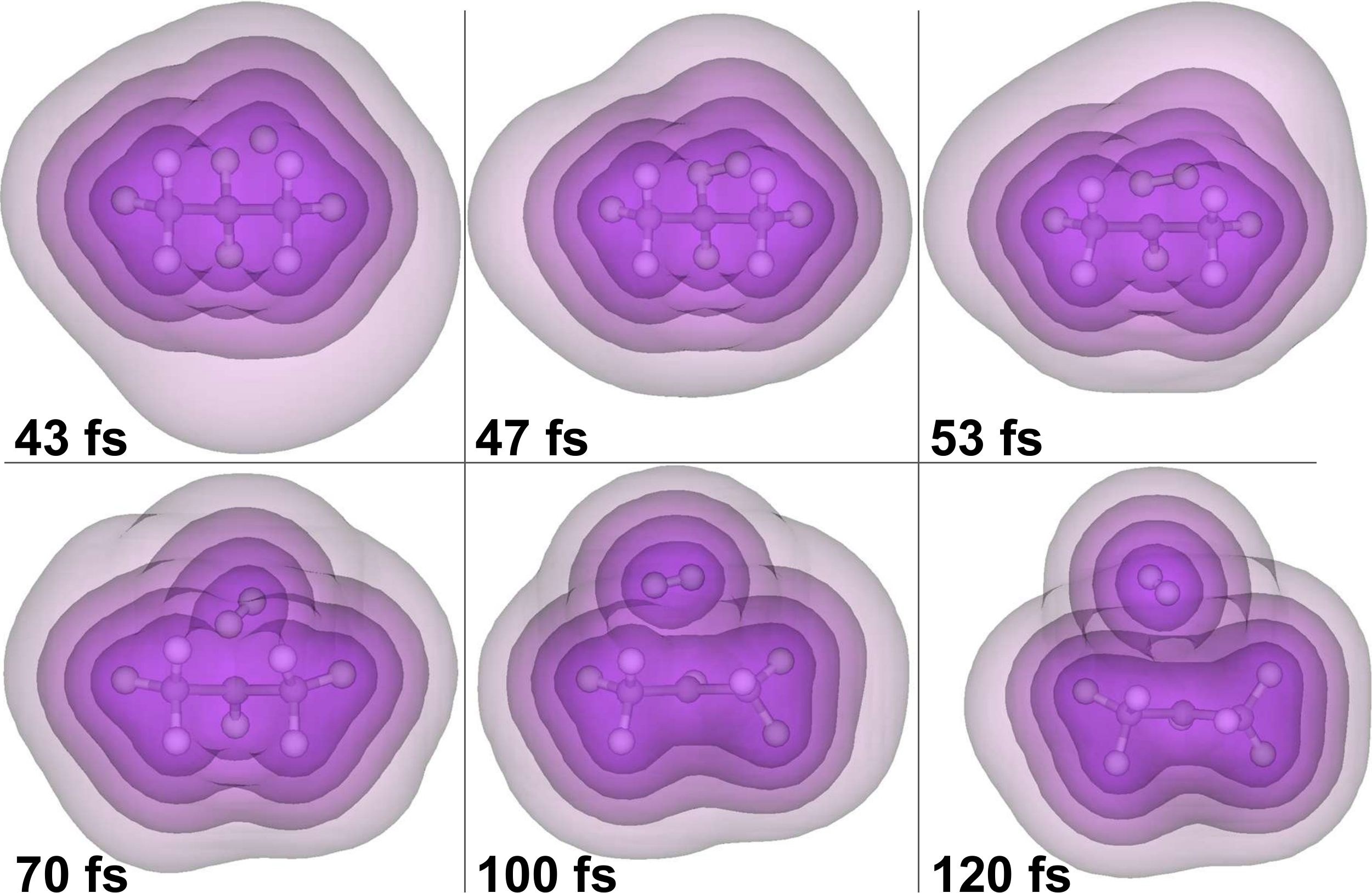}
\caption{Abstraction dynamics of a proton incident on C\textsubscript{3}H\textsubscript{8}, directed toward IP~1 with an initial KE of 0.52~eV. The electron density isosurfaces are shown in purple at values of 0.5, 0.1, 0.001, and 0.0001.}
\label{fig:c3h8-ip2-v-0.1}
\end{figure*}

Another noteworthy case in Table~\ref{tab:c3h8-data} is the formation of an H\textsubscript{2} molecule at IP~2, associated with an initial proton kinetic energy of 0.52~eV. The dynamics of this process are shown in Fig.~\ref{fig:c3h8-ip2-v-0.1}. By 43~fs, the incident proton has approached the molecule, and at 47~fs it comes sufficiently close to transiently bond with the top-center H atom of C\textsubscript{3}H\textsubscript{8}. By 53~fs, the proton and the bonded H atom detach from the parent molecule and are repelled upward, away from the C\textsubscript{3}H\textsubscript{7} fragment, reaching complete separation by 120~fs. This result demonstrates that proton collisions can induce the formation of H\textsubscript{2}, underscoring the complex dynamics of these interactions and the ability of protons to abstract molecular fragments. More broadly, it highlights the role of many-body effects in Coulomb explosion and suggests that a wider variety of fragmentation products may emerge through such mechanisms~\cite{PhysRevA.111.013109,Mogyorosi2025}.

Interestingly, in terms of reaction type, Table~\ref{tab:c3h8-data} shows a striking resemblance to Table~\ref{tab:c2h2-data}. In both molecules, IP~1 results exclusively in scattering outcomes. This is consistent with the geometry, as IP~1 corresponds to the center of the molecule—directly at C--C bonds or central C atoms (see Fig.~\ref{fig:c2h2-ip-diagram} and Fig.~\ref{fig:c3h8-ip-diagram}). In this configuration, there is insufficient space for the proton to form a bond, and it is repelled by the carbon nuclear core.  

For the remaining incident points corresponding to the right C atom, the right C–H bond, and the right H atom (IP~2–4 in C\textsubscript{2}H\textsubscript{2} and IP~3–5 in C\textsubscript{3}H\textsubscript{8}), proton capture occurs at the lowest KE of 0.52~eV. This strong correlation reflects the similarity in the collision geometry at these sites (see Fig.~\ref{fig:c2h2-ip-diagram} and Fig.~\ref{fig:c3h8-ip-diagram}). At higher proton energies, the reaction patterns are also analogous between the two molecules. For example, the row of reaction types at IP~2 in C\textsubscript{2}H\textsubscript{2} matches that of IP~3 in C\textsubscript{3}H\textsubscript{8}\textemdash proton capture at the lowest KE followed by scattering at higher KEs. Similarly, IP~3 in C\textsubscript{2}H\textsubscript{2} and IP~4 in C\textsubscript{3}H\textsubscript{8} exhibit scattering at all KEs except for a specific critical energy range, where abstraction occurs, as discussed in Sec.~\ref{subsec:acetylene} (25.39~eV and 12.96~eV for C\textsubscript{2}H\textsubscript{2} and C\textsubscript{3}H\textsubscript{8}, respectively).  

Finally, the right-most incident point in both molecules (IP~4 for C\textsubscript{2}H\textsubscript{2} and IP~5 for C\textsubscript{3}H\textsubscript{8}) consistently results in abstraction at all KEs above the lowest tested value. This highlights that striking the terminal hydrogen in a hydrocarbon leads to a high probability of abstraction, as it allows the proton to efficiently transfer nearly all its KE to the target H atom, with the residual energy redistributed throughout the hydrocarbon molecule.

\subsection{Butane (C\textsubscript{4}H\textsubscript{10})}
\label{subsec:butane}

\begin{figure}[ht!]
    \centering
    \includegraphics[width=0.95\columnwidth]{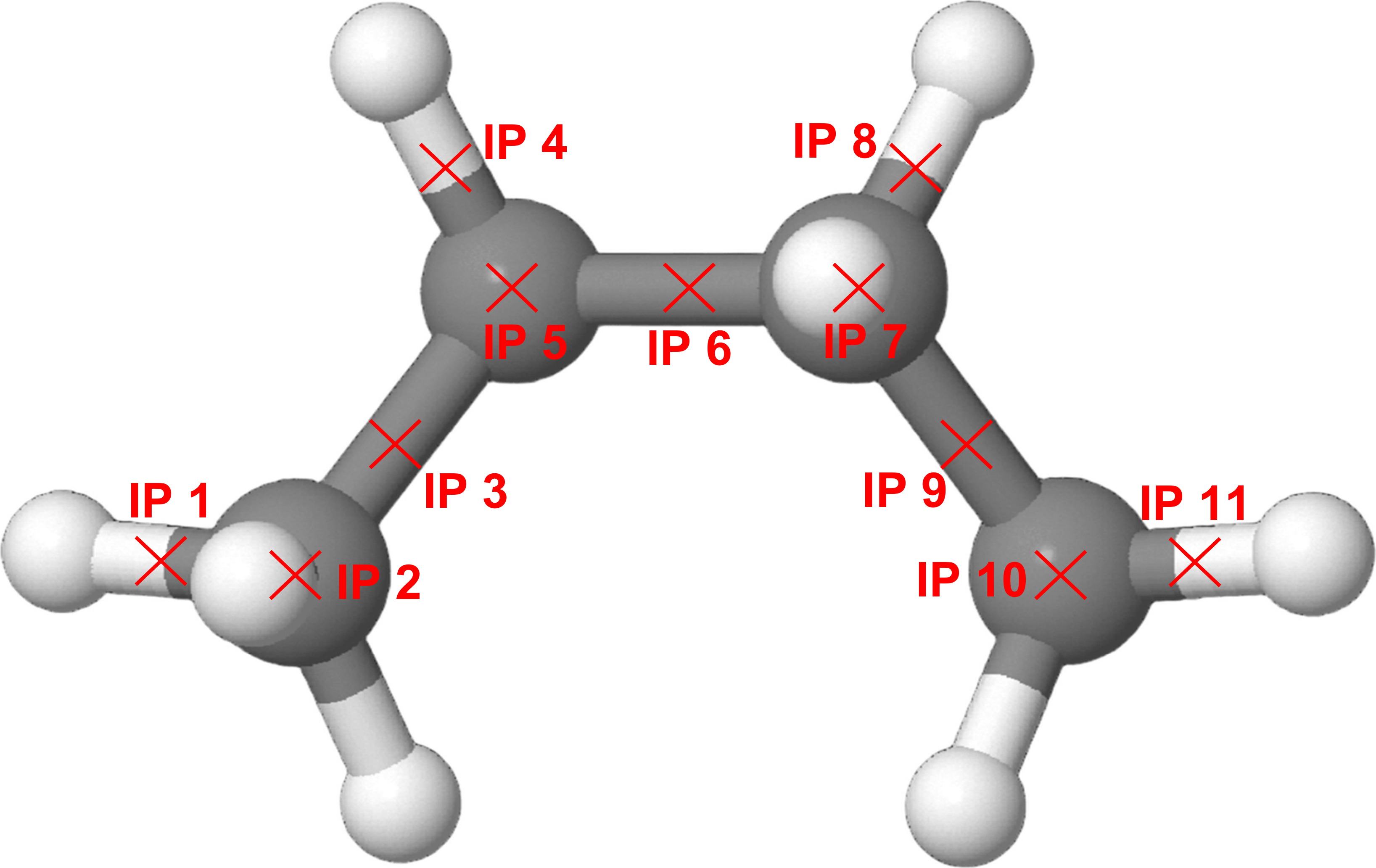}
    \caption{Impact point (IP) diagram for C\textsubscript{4}H\textsubscript{10}.}
    \label{fig:c4h10-ip-diagram}
\end{figure}
For the butane molecule, we selected the gauche conformation. This
structural choice was made to increase diversity, as it
offers a greater number of potential impact points and enables the
investigation of a non-centrosymmetric molecular system.
The selected IPs for C\textsubscript{4}H\textsubscript{10} are shown in Fig.~\ref{fig:c4h10-ip-diagram}. 
As in the cases of C\textsubscript{2}H\textsubscript{2} and C\textsubscript{3}H\textsubscript{8} 
(discussed in Sec.~\ref{subsec:acetylene} and Sec.~\ref{subsec:propane}, respectively), 
the IPs were chosen to represent regions of interest on the molecular framework, 
specifically bond centers and atomic sites. In contrast to C\textsubscript{2}H\textsubscript{2} 
and C\textsubscript{3}H\textsubscript{8}, however, C\textsubscript{4}H\textsubscript{10} 
lacks left–right symmetry, requiring the selection of IPs on both sides of 
the molecule (see Fig.~\ref{fig:c4h10-ip-diagram}).

\begin{table*}[t]
\centering
\renewcommand{\arraystretch}{1.2}
\setcellgapes{2pt}\makegapedcells
\begin{tabular}{lccccccc}
\toprule
\makecell{\textbf{Impact Point (IP)}\\\textit{Proton Kinetic Energy (eV)}} & \textbf{0.52} & \textbf{4.66} & \textbf{12.96} & \textbf{25.39} & \textbf{41.98} & \textbf{62.70} & \textbf{87.58} \\
\midrule
\arrayrulecolor[gray]{0.75}
\textbf{IP 1} & \makecell{P, -- \\ $\mathrm{C_4H_{11}}^{1.23+}$} & \makecell{P, -- \\ $\mathrm{C_4H_{11}}^{1.25+}$} & \makecell{S, 6.93 \\ $\mathrm{C_4H_{10}}^{0.93+}$ \\ $\mathrm{H}^{0.36+}$} & \makecell{S, 16.05 \\ $\mathrm{C_4H_{9}}^{1.00+}$ \\ $\mathrm{H}^{0.11+}$ \\ $\mathrm{H}^{0.22+}$} & \makecell{S, 39.45 \\ $\mathrm{C_4H_{9}}^{0.97+}$ \\ $\mathrm{H}^{0.25+}$ \\ $\mathrm{H}^{0.14+}$} & \makecell{A, -- \\ $\mathrm{C_4H_{10}}^{1.05+}$ \\ $\mathrm{H}^{0.33+}$} & \makecell{S, 47.61 \\ $\mathrm{C_4H_{9}}^{0.91+}$ \\ $\mathrm{H}^{0.27+}$ \\ $\mathrm{H}^{0.15+}$} \\
\midrule
\textbf{IP 2} & \makecell{A, -- \\ $\mathrm{C_4H_{9}}^{1.19+}$ \\ $\mathrm{H_2}^{0.01+}$} & \makecell{A, -- \\ $\mathrm{C_4H_{9}}^{1.24+}$ \\ $\mathrm{H_2}^{0.06+}$} & \makecell{A, -- \\ $\mathrm{C_4H_{10}}^{1.14+}$ \\ $\mathrm{H}^{0.20+}$} & \makecell{S, 23.98 \\ $\mathrm{C_4H_{9}}^{0.92+}$ \\ $\mathrm{H}^{0.12+}$ \\ $\mathrm{H}^{0.40+}$} & \makecell{S, 37.12 \\ $\mathrm{C_4H_{9}}^{0.92+}$ \\ $\mathrm{H}^{0.44+}$ \\ $\mathrm{H}^{0.12+}$} & \makecell{S, 41.13 \\ $\mathrm{C_4H_{9}}^{0.93+}$ \\ $\mathrm{H}^{0.47+}$ \\ $\mathrm{H}^{0.05+}$} & \makecell{S, 44.35 \\ $\mathrm{C_4H_{9}}^{0.72+}$ \\ $\mathrm{H}^{0.42+}$ \\ $\mathrm{H}^{0.31+}$} \\
\midrule
\textbf{IP 3} & \makecell{P, -- \\ $\mathrm{C_4H_{11}}^{1.22+}$} & \makecell{P, -- \\ $\mathrm{C_4H_{11}}^{1.33+}$} & \makecell{S, 11.76 \\ $\mathrm{CH_3}^{0.28+}$ \\ $\mathrm{C_3H_7}^{0.84+}$ \\ $\mathrm{H}^{0.24+}$} & \makecell{S, 12.85 \\ $\mathrm{C_4H_{10}}^{1.07+}$ \\ $\mathrm{H}^{1.00+}$} & \makecell{S, 9.30 \\ $\mathrm{C_4H_{10}}^{1.01+}$ \\ $\mathrm{H}^{0.38+}$} & \makecell{S, 8.80 \\ $\mathrm{C_4H_{10}}^{1.02+}$ \\ $\mathrm{H}^{0.41+}$} & \makecell{S, 9.19 \\ $\mathrm{C_4H_{10}}^{1.06+}$ \\ $\mathrm{H}^{0.42+}$} \\
\midrule
\textbf{IP 4} & \makecell{S, 0.37 \\ $\mathrm{C_4H_{9}}^{1.13+}$ \\ $\mathrm{H}^{0.02-}$ \\ $\mathrm{H}^{0.09+}$} & \makecell{S, 3.58 \\ $\mathrm{C_4H_{10}}^{0.78+}$ \\ $\mathrm{H}^{0.49+}$} & \makecell{A, -- \\ $\mathrm{C_4H_{9}}^{1.27+}$ \\ $\mathrm{H_2}^{0.06+}$} & \makecell{S, 12.71 \\ $\mathrm{C_4H_{9}}^{1.15+}$ \\ $\mathrm{H}^{0.08+}$ \\ $\mathrm{H}^{0.14+}$} & \makecell{S, 11.07 \\ $\mathrm{C_4H_{9}}^{1.01+}$ \\ $\mathrm{H}^{0.08+}$ \\ $\mathrm{H}^{0.27+}$} & \makecell{S, 10.31 \\ $\mathrm{C_4H_{10}}^{0.94+}$ \\ $\mathrm{H}^{0.47+}$} & \makecell{S, 9.67 \\ $\mathrm{C_4H_{10}}^{0.94+}$ \\ $\mathrm{H}^{0.44+}$} \\
\midrule
\textbf{IP 5} & \makecell{S, 0.36 \\ $\mathrm{C_4H_{9}}^{1.16+}$ \\ $\mathrm{H}^{0.03+}$ \\ $\mathrm{H}^{0.05+}$} & \makecell{S, 3.44 \\ $\mathrm{C_4H_{10}}^{0.86+}$ \\ $\mathrm{H}^{0.45+}$} & \makecell{S, 4.95 \\ $\mathrm{C_4H_{10}}^{0.88+}$ \\ $\mathrm{H}^{0.39+}$} & \makecell{S, 9.81 \\ $\mathrm{C_4H_{10}}^{0.86+}$ \\ $\mathrm{H}^{0.48+}$} & \makecell{S, 16.03 \\ $\mathrm{C_4H_{10}}^{1.01+}$ \\ $\mathrm{H}^{0.45+}$} & \makecell{S, 22.33 \\ $\mathrm{C_4H_{10}}^{1.06+}$ \\ $\mathrm{H}^{0.39+}$} & \makecell{S, 30.32 \\ $\mathrm{C_2H_5}^{0.51+}$ \\ $\mathrm{C_2H_5}^{0.51+}$ \\ $\mathrm{H}^{0.45+}$} \\
\midrule
\textbf{IP 6} & \makecell{P, -- \\ $\mathrm{C_4H_{11}}^{1.24+}$} & \makecell{A, -- \\ $\mathrm{C_4H_{9}}^{1.27+}$ \\ $\mathrm{H_2}^{0.05+}$} & \makecell{S, 7.10 \\ $\mathrm{C_4H_{10}}^{0.94+}$ \\ $\mathrm{H}^{0.43+}$} & \makecell{S, 16.53 \\ $\mathrm{C_2H_5}^{0.59+}$ \\ $\mathrm{C_2H_5}^{0.62+}$ \\ $\mathrm{H}^{0.23+}$} & \makecell{S, 14.30 \\ $\mathrm{C_2H_5}^{0.70+}$ \\ $\mathrm{C_2H_5}^{0.51+}$ \\ $\mathrm{H}^{0.29+}$} & \makecell{S, 11.04 \\ $\mathrm{C_4H_{10}}^{1.08+}$ \\ $\mathrm{H}^{0.37+}$} & \makecell{S, 10.18 \\ $\mathrm{C_4H_{10}}^{1.08+}$ \\ $\mathrm{H}^{0.38+}$} \\
\midrule
\textbf{IP 7} & \makecell{P, -- \\ $\mathrm{C_4H_{11}}^{1.25+}$} & \makecell{A, -- \\ $\mathrm{C_4H_{9}}^{1.24+}$ \\ $\mathrm{H_2}^{0.05+}$} & \makecell{A, -- \\ $\mathrm{C_4H_{10}}^{0.90+}$ \\ $\mathrm{H}^{0.48+}$} & \makecell{S, 21.65 \\ $\mathrm{C_4H_{9}}^{1.19+}$ \\ $\mathrm{H}^{0.09+}$ \\ $\mathrm{H}^{0.07+}$} & \makecell{S, 34.43 \\ $\mathrm{C_4H_{9}}^{1.03+}$ \\ $\mathrm{H}^{0.30+}$ \\ $\mathrm{H}^{0.07+}$} & \makecell{S, 49.35 \\ $\mathrm{C_4H_{9}}^{0.78+}$ \\ $\mathrm{H}^{0.42+}$ \\ $\mathrm{H}^{0.37+}$} & \makecell{S, 55.73 \\ $\mathrm{C_4H_{9}}^{0.74+}$ \\ $\mathrm{H}^{0.36+}$ \\ $\mathrm{H}^{0.37+}$} \\
\midrule
\textbf{IP 8} & \makecell{P, -- \\ $\mathrm{C_4H_{11}}^{1.23+}$} & \makecell{S, 4.31 \\ $\mathrm{C_4H_{9}}^{1.24+}$ \\ $\mathrm{H}^{0.01+}$ \\ $\mathrm{H}^{0.07+}$} & \makecell{A, -- \\ $\mathrm{C_4H_{9}}^{1.16+}$ \\ $\mathrm{H_2}^{0.17+}$} & \makecell{S, 22.67 \\ $\mathrm{C_4H_{9}}^{1.14+}$ \\ $\mathrm{H}^{0.19+}$ \\ $\mathrm{H}^{0.08+}$} & \makecell{S, 23.23 \\ $\mathrm{C_4H_{9}}^{1.13+}$ \\ $\mathrm{H}^{0.16+}$ \\ $\mathrm{H}^{0.07+}$} & \makecell{S, 16.89 \\ $\mathrm{C_4H_{9}}^{1.09+}$ \\ $\mathrm{H}^{0.08+}$ \\ $\mathrm{H}^{0.14+}$} & \makecell{S, 14.65 \\ $\mathrm{C_4H_{9}}^{1.04+}$ \\ $\mathrm{H}^{0.09+}$ \\ $\mathrm{H}^{0.25+}$} \\
\midrule
\textbf{IP 9} & \makecell{S, 0.44 \\ $\mathrm{C_4H_{9}}^{1.18+}$ \\ $\mathrm{H}^{0.08+}$ \\ $\mathrm{H}^{0.01-}$} & \makecell{S, 4.52 \\ $\mathrm{C_4H_{9}}^{1.25+}$ \\ $\mathrm{H}^{0.03+}$ \\ $\mathrm{H}^{0.01+}$} & \makecell{S, 12.06 \\ $\mathrm{C_3H_7}^{0.75+}$ \\ $\mathrm{CH_3}^{0.39+}$ \\ $\mathrm{H}^{0.21+}$} & \makecell{S, 11.52 \\ $\mathrm{C_4H_{10}}^{0.95+}$ \\ $\mathrm{H}^{0.48+}$} & \makecell{S, 9.22 \\ $\mathrm{C_4H_{10}}^{1.10+}$ \\ $\mathrm{H}^{0.34+}$} & \makecell{S, 9.31 \\ $\mathrm{C_4H_{10}}^{1.07+}$ \\ $\mathrm{H}^{0.40+}$} & \makecell{S, 8.79 \\ $\mathrm{C_4H_{10}}^{1.05+}$ \\ $\mathrm{H}^{0.32+}$} \\
\midrule
\textbf{IP 10} & \makecell{P, -- \\ $\mathrm{C_4H_{11}}^{1.25+}$} & \makecell{S, 2.62 \\ $\mathrm{C_4H_{10}}^{0.84+}$ \\ $\mathrm{H}^{0.42+}$} & \makecell{S, 7.11 \\ $\mathrm{C_4H_{10}}^{0.98+}$ \\ $\mathrm{H}^{0.37+}$} & \makecell{S, 10.23 \\ $\mathrm{C_4H_{10}}^{1.01+}$ \\ $\mathrm{H}^{0.38+}$} & \makecell{S, 15.07 \\ $\mathrm{C_3H_7}^{0.81+}$ \\ $\mathrm{CH_3}^{0.27+}$ \\ $\mathrm{H}^{0.35+}$} & \makecell{S, 21.51 \\ $\mathrm{C_3H_7}^{0.74+}$ \\ $\mathrm{CH_3}^{0.25+}$ \\ $\mathrm{H}^{0.41+}$} & \makecell{S, 29.04 \\ $\mathrm{C_3H_7}^{0.65+}$ \\ $\mathrm{CH_3}^{0.34+}$ \\ $\mathrm{H}^{0.40+}$} \\
\midrule
\textbf{IP 11} & \makecell{P, -- \\ $\mathrm{C_4H_{11}}^{1.26+}$} & \makecell{A, -- \\ $\mathrm{C_4H_{10}}^{1.31+}$ \\ $\mathrm{H}^{0.04+}$} & \makecell{A, -- \\ $\mathrm{C_4H_{9}}^{1.16+}$ \\ $\mathrm{H_2}^{0.14+}$} & \makecell{A, -- \\ $\mathrm{C_4H_{9}}^{1.03+}$ \\ $\mathrm{H_2}^{0.19+}$} & \makecell{S, 12.45 \\ $\mathrm{C_4H_{9}}^{0.97+}$ \\ $\mathrm{H}^{0.15+}$ \\ $\mathrm{H}^{0.29+}$} & \makecell{S, 11.03 \\ $\mathrm{C_4H_{10}}^{1.06+}$ \\ $\mathrm{H}^{0.36+}$} & \makecell{S, 10.60 \\ $\mathrm{C_4H_{10}}^{1.02+}$ \\ $\mathrm{H}^{0.43+}$} \\
\arrayrulecolor{black}
\bottomrule
\end{tabular}
\caption{Combined outcome data for different C\textsubscript{4}H\textsubscript{10} IPs (rows) and proton KEs (columns). Each cell shows the reaction type (S: scattering, P: proton capture, A: abstraction), followed by the KE loss of the proton and the resulting fragment products with their respective charges.}
\label{tab:c4h10-data}
\end{table*}

The results of the simulations, including reaction type, KE loss, and fragmentation products, are summarized in Table~\ref{tab:c4h10-data}. The data highlight the dependence of the outcomes on both the location of the proton collision and the KE of the incoming projectile. At the highest tested KE of 87.58~eV, protons scatter regardless of the chosen IP. In contrast, at the lowest tested KE of 0.52~eV, proton capture is the most frequent outcome, occurring in seven cases across the various IPs. The specific IP also plays a significant role in determining the reaction pathway; for example, IP~5 and IP~9 consistently lead to scattering, independent of the proton projectile KE.

\begin{figure*}
\centering
\includegraphics[width=\textwidth]{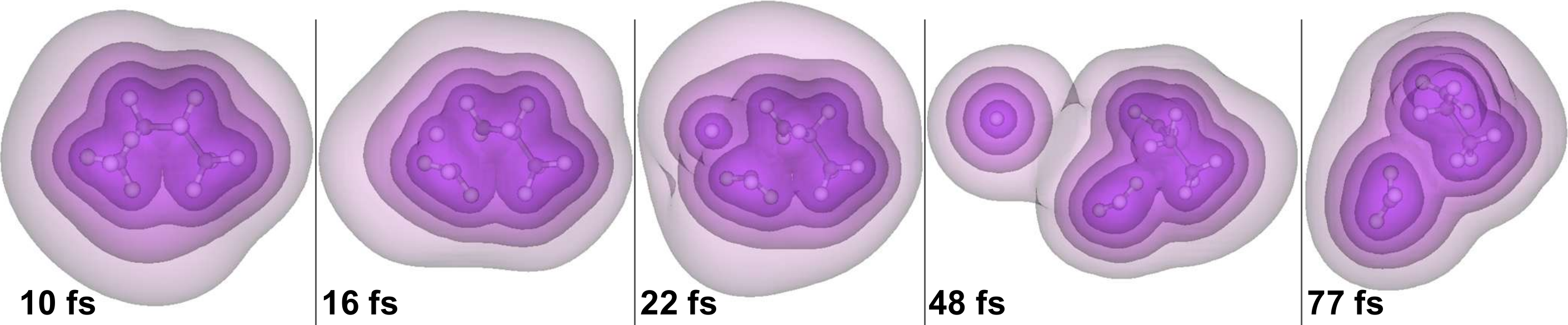}
\caption{Scattering dynamics of a proton incident on C\textsubscript{4}H\textsubscript{10}, directed toward IP~3 with an initial KE of 12.96~eV. The electron density isosurfaces are shown in purple at values of 0.5, 0.1, 0.001, and 0.0001.}
\label{fig:c4h10-ip3-v-0.5}
\end{figure*}

The data also reveals diverse fragmentation pathways resulting from proton collisions with C\textsubscript{4}H\textsubscript{10}. For example, fragments such as C\textsubscript{4}H\textsubscript{9}, C\textsubscript{3}H\textsubscript{7}, C\textsubscript{2}H\textsubscript{5}, and CH\textsubscript{3} are observed across different IPs and proton KEs. Many of these fragmentation events occur only under very specific conditions. A notable case arises at IP~3 with a proton KE of 12.96~eV, where the collision cleaves the molecule into CH\textsubscript{3} and C\textsubscript{3}H\textsubscript{7} fragments. At higher KEs, no fragmentation occurs at this IP, while at lower KEs, the molecule instead captures the proton. This demonstrates that such fragmentation events are uncommon and require narrowly defined conditions.

The molecular dynamics of the IP~3, 12.96~eV case are illustrated in Fig.~\ref{fig:c4h10-ip3-v-0.5}. As shown, the proton reaches the center of the left C--C bond at 10~fs. By 16~fs, the proton repels the positively charged carbon nuclei, leading to bond cleavage into CH\textsubscript{3} and C\textsubscript{3}H\textsubscript{7}, while the proton itself is repelled and ejected from the molecular core. From 22~fs to 77~fs, the proton continues migrating away as the two fragments separate and undergo internal rearrangement, ultimately forming the distinct CH\textsubscript{3} and C\textsubscript{3}H\textsubscript{7} products. This trajectory highlights why the fragmentation and CH\textsubscript{3} formation probability is low: successful breakup requires an optimal set of conditions in which the proton KE is sufficient to penetrate the C--C bond center without being displaced beforehand.

At lower KEs (4.66~eV and below), the proton lacks sufficient energy to overcome the potential barrier and reach the C--C bond. In these cases, charge redistribution occurs over a longer timescale, leading to proton capture rather than bond cleavage. At higher KEs (25.39~eV and above), the proton traverses the system too quickly, transferring excess energy and scattering past the target bond rather than inducing fragmentation. Thus, fragmentation occurs only within a narrow KE window near 12.96~eV. This optimal energy range for cleaving terminal C--C bonds is consistent with the results obtained for propane (Sec.~\ref{subsec:propane}). Specifically, IP~3/9 in Table~\ref{tab:c4h10-data} and IP~2 in Table~\ref{tab:c3h8-data} correspond to terminal C--C bonds in C\textsubscript{4}H\textsubscript{10} and C\textsubscript{3}H\textsubscript{8}, respectively. In both molecules, collisions at these sites with a KE of 12.96~eV lead to the separation of a terminal CH\textsubscript{3} group from the parent molecule. These findings underscore the similarities across different alkanes and suggest that proton impact energies in this range are optimal for cleaving terminal C--C bonds.

The energy required to cleave the central C--C bond (IP~6 in Fig.~\ref{fig:c4h10-ip-diagram}) is greater than that needed to break the terminal C--C bond discussed above. As shown in Table~\ref{tab:c4h10-data} at IP~6, proton KEs of 25.39 and 41.98~eV result in cleavage of the central bond, separating C\textsubscript{4}H\textsubscript{10} into two equal C\textsubscript{2}H\textsubscript{5} fragments. At both lower and higher KEs, however, this bond remains intact.

In contrast to the case where CH\textsubscript{3} and C\textsubscript{3}H\textsubscript{7} fragments are produced only within a restricted KE range by directly striking the terminal C--C bond (IP~3/9), fragmentation is more robust when the terminal C atom is impacted directly, as in IP~10 (see Fig.~\ref{fig:c4h10-ip-diagram}). As shown in Table~\ref{tab:c4h10-data}, every proton KE at or above 41.98~eV for IP~10 results in C--C bond cleavage and the subsequent formation of CH\textsubscript{3} and C\textsubscript{3}H\textsubscript{7}. In this case, the proton carries sufficient energy to collide with the terminal carbon nucleus and eject it from the molecular framework.

An important aspect of the fragmentation results concerns the charge states of the resulting fragments. 
The fractional charge states, obtained by integrating the electron density using TDDFT with 
the ALDA functional, can be interpreted as probabilities of discrete charge states. For instance, 
a fragment identified as CH\textsubscript{3}$^{0.28+}$ (see Table~\ref{tab:c4h10-data} for 12.96~eV at IP~3) 
corresponds to a 28\% probability of being singly charged and a 72\% probability of being neutral. 
These results therefore predict the formation of neutral as well as charged fragments, offering 
insight into their role in Coulomb explosion experiments. This is particularly significant because 
the neutral counterparts of odd-electron fragments correspond to radical species, which have been 
highlighted as a subject of interest in the Coulomb explosion of butane~\cite{Mogyorosi2025}.
One could alternatively contend that the fractional charges observed
in molecular fragments arise from computational
limitations—specifically, the constrained simulation box size and
finite simulation duration. Under this interpretation, the electron
cloud disrupted by proton scattering would eventually redistribute to
yield integer charges if given adequate time to equilibrate within a
sufficiently expansive simulation domain.

\section{Summary}

We have used time-dependent density functional theory to investigate low-energy proton collisions with hydrocarbons of increasing complexity: acetylene (C\textsubscript{2}H\textsubscript{2}), propane (C\textsubscript{3}H\textsubscript{8}), and butane (C\textsubscript{4}H\textsubscript{10}). By systematically varying both the incident point (IP) and the initial proton kinetic energy (KE), we identified general trends in the reaction outcomes as well as molecule-specific fragmentation pathways.

Across all systems, three principal reaction types were observed: scattering, proton capture, and abstraction. The outcome depends sensitively on both the proton KE and the collision geometry (See Tables~\ref{tab:c2h2-data},~\ref{tab:c3h8-data},~and~\ref{tab:c4h10-data} for C\textsubscript{2}H\textsubscript{2}, C\textsubscript{3}H\textsubscript{8}, and C\textsubscript{4}H\textsubscript{10} results, respectively). IPs at central C--C bonds and C atoms predominantly lead to scattering, while terminal H atoms consistently favor abstraction at KEs above a few electronvolts. At the lowest tested energy (0.52~eV), proton capture dominates at most IPs, where the slow approach enables charge redistribution and local atomic rearrangements that stabilize the projectile within the hydrocarbon framework.

C--C bond cleavage was found to be uncommon and sensitive to initial conditions. A narrow KE window near 12.96~eV was identified as optimal for detaching a terminal CH\textsubscript{3} group in both propane and butane when striking at the C--C bond. At lower energies, the proton is captured before reaching the bond, while at higher energies, it traverses too quickly and scatters. These results reveal that intermediate KEs are uniquely effective at inducing fragmentation and formation of smaller hydrocarbons as many different fragment products have been observed across various KEs and IPs such as C\textsubscript{4}H\textsubscript{9}, C\textsubscript{3}H\textsubscript{7}, C\textsubscript{2}H\textsubscript{5}, and CH\textsubscript{3} stemming from the parent molecule of C\textsubscript{4}H\textsubscript{10}. Interestingly, the snapshots reveal complicated dynamics at play such as hydrogen migration and hydrogen abstraction leading to unique fragments in C\textsubscript{3}H\textsubscript{8} such as H\textsubscript{2}, C\textsubscript{2}H\textsubscript{5}, CH\textsubscript{2}, and CH\textsubscript{3}. Such outcomes underscore the importance of including proton–molecule collisions when modeling fragment distributions and Coulomb explosion interactions.

Charge-state analysis showed that fragments frequently carry fractional charges, which can be interpreted as probabilities of discrete charge states. This indicates that neutral fragments, including radical species such as CH\textsubscript{3}, can be produced alongside ionic products. Such radicals have been reported in experimental studies of hydrocarbon Coulomb explosion, and our results provide an additional possible microscopic mechanism for their formation~\cite{Mogyorosi2025}.

Overall, our findings demonstrate that fragmentation dynamics in hydrocarbons under proton impact arise from a delicate interplay of projectile energy and molecular geometry. The consistent trends observed across acetylene, propane, and butane point toward general features of alkanes such that: C--C bonds are susceptible to breakup only within a narrow intermediate KE range, while terminal hydrogens favor abstraction across a broad range of energies. In the context of Coulomb explosion, these results highlight the role of secondary proton–molecule collisions in generating a wide variety of fragments beyond those directly produced by laser-induced ionization. More broadly, they contribute to a microscopic understanding of ion–molecule interactions relevant to radiation damage, ion-beam processing, and astrochemical environments where low-energy proton collisions play a central role.

Future work could involve experimental validation of these results as well as extending the calculations to larger molecules, to collisions involving other ionic projectiles, to finite-temperature conditions, and to trajectories in which the projectiles impinge at different angles. Such studies would further clarify the generality of the trends identified here and help establish connections to real experimental conditions.

\begin{acknowledgments} 
This work has been supported by the National Science Foundation (NSF)
under Grants No. DMR-2217759 and No. NSF IRES 2245029.
This work used ACES at TAMU through allocation PHYS240167 from the Advanced Cyberinfrastructure Coordination Ecosystem: Services \& Support (ACCESS) program, which is supported by National Science 
Foundation grants \#2138259, \#2138286, \#2138307, \#2137603, and \#2138296~\cite{aces}.

The ELI ALPS project (GINOP-2.3.6-15-2015-00001) is supported by the European Union and co-financed by the European Regional Development Fund.

\end{acknowledgments}


\begin{thebibliography}{63}%
\makeatletter
\providecommand \@ifxundefined [1]{%
 \@ifx{#1\undefined}
}%
\providecommand \@ifnum [1]{%
 \ifnum #1\expandafter \@firstoftwo
 \else \expandafter \@secondoftwo
 \fi
}%
\providecommand \@ifx [1]{%
 \ifx #1\expandafter \@firstoftwo
 \else \expandafter \@secondoftwo
 \fi
}%
\providecommand \natexlab [1]{#1}%
\providecommand \enquote  [1]{``#1''}%
\providecommand \bibnamefont  [1]{#1}%
\providecommand \bibfnamefont [1]{#1}%
\providecommand \citenamefont [1]{#1}%
\providecommand \href@noop [0]{\@secondoftwo}%
\providecommand \href [0]{\begingroup \@sanitize@url \@href}%
\providecommand \@href[1]{\@@startlink{#1}\@@href}%
\providecommand \@@href[1]{\endgroup#1\@@endlink}%
\providecommand \@sanitize@url [0]{\catcode `\\12\catcode `\$12\catcode
  `\&12\catcode `\#12\catcode `\^12\catcode `\_12\catcode `\%12\relax}%
\providecommand \@@startlink[1]{}%
\providecommand \@@endlink[0]{}%
\providecommand \url  [0]{\begingroup\@sanitize@url \@url }%
\providecommand \@url [1]{\endgroup\@href {#1}{\urlprefix }}%
\providecommand \urlprefix  [0]{URL }%
\providecommand \Eprint [0]{\href }%
\providecommand \doibase [0]{https://doi.org/}%
\providecommand \selectlanguage [0]{\@gobble}%
\providecommand \bibinfo  [0]{\@secondoftwo}%
\providecommand \bibfield  [0]{\@secondoftwo}%
\providecommand \translation [1]{[#1]}%
\providecommand \BibitemOpen [0]{}%
\providecommand \bibitemStop [0]{}%
\providecommand \bibitemNoStop [0]{.\EOS\space}%
\providecommand \EOS [0]{\spacefactor3000\relax}%
\providecommand \BibitemShut  [1]{\csname bibitem#1\endcsname}%
\let\auto@bib@innerbib\@empty
\bibitem [{\citenamefont {Carante}\ \emph {et~al.}(2024)\citenamefont
  {Carante}, \citenamefont {Ramos},\ and\ \citenamefont
  {Ballarini}}]{ijms25126368}%
  \BibitemOpen
  \bibfield  {author} {\bibinfo {author} {\bibfnamefont {M.~P.}\ \bibnamefont
  {Carante}}, \bibinfo {author} {\bibfnamefont {R.~L.}\ \bibnamefont {Ramos}},\
  and\ \bibinfo {author} {\bibfnamefont {F.}~\bibnamefont {Ballarini}},\
  }\bibfield  {title} {\bibinfo {title} {Radiation damage in biomolecules and
  cells 3.0},\ }\bibfield  {journal} {\bibinfo  {journal} {International
  Journal of Molecular Sciences}\ }\textbf {\bibinfo {volume} {25}},\ \href
  {https://doi.org/10.3390/ijms25126368} {10.3390/ijms25126368} (\bibinfo
  {year} {2024})\BibitemShut {NoStop}%
\bibitem [{\citenamefont {Shepard}\ \emph {et~al.}(2023)\citenamefont
  {Shepard}, \citenamefont {Yost},\ and\ \citenamefont
  {Kanai}}]{PhysRevLett.130.118401}%
  \BibitemOpen
  \bibfield  {author} {\bibinfo {author} {\bibfnamefont {C.}~\bibnamefont
  {Shepard}}, \bibinfo {author} {\bibfnamefont {D.~C.}\ \bibnamefont {Yost}},\
  and\ \bibinfo {author} {\bibfnamefont {Y.}~\bibnamefont {Kanai}},\ }\bibfield
   {title} {\bibinfo {title} {Electronic excitation response of dna to
  high-energy proton radiation in water},\ }\href
  {https://doi.org/10.1103/PhysRevLett.130.118401} {\bibfield  {journal}
  {\bibinfo  {journal} {Phys. Rev. Lett.}\ }\textbf {\bibinfo {volume} {130}},\
  \bibinfo {pages} {118401} (\bibinfo {year} {2023})}\BibitemShut {NoStop}%
\bibitem [{\citenamefont {KC}\ and\ \citenamefont {Abolfath}(2022)}]{KC2022}%
  \BibitemOpen
  \bibfield  {author} {\bibinfo {author} {\bibfnamefont {S.}~\bibnamefont
  {KC}}\ and\ \bibinfo {author} {\bibfnamefont {R.}~\bibnamefont {Abolfath}},\
  }\bibfield  {title} {\bibinfo {title} {Towards the ionizing radiation induced
  bond dissociation mechanism in oxygen, water, guanine and dna fragmentation:
  a density functional theory simulation},\ }\href
  {https://doi.org/10.1038/s41598-022-23727-3} {\bibfield  {journal} {\bibinfo
  {journal} {Scientific Reports}\ }\textbf {\bibinfo {volume} {12}},\ \bibinfo
  {pages} {19853} (\bibinfo {year} {2022})}\BibitemShut {NoStop}%
\bibitem [{\citenamefont {Saha}\ \emph {et~al.}(2024)\citenamefont {Saha},
  \citenamefont {Mecklenburg}, \citenamefont {Pattison}, \citenamefont
  {Brewster}, \citenamefont {Rodriguez},\ and\ \citenamefont
  {Ercius}}]{10.1093/mam/ozae044.902}%
  \BibitemOpen
  \bibfield  {author} {\bibinfo {author} {\bibfnamefont {A.}~\bibnamefont
  {Saha}}, \bibinfo {author} {\bibfnamefont {M.}~\bibnamefont {Mecklenburg}},
  \bibinfo {author} {\bibfnamefont {A.}~\bibnamefont {Pattison}}, \bibinfo
  {author} {\bibfnamefont {A.}~\bibnamefont {Brewster}}, \bibinfo {author}
  {\bibfnamefont {J.~A.}\ \bibnamefont {Rodriguez}},\ and\ \bibinfo {author}
  {\bibfnamefont {P.}~\bibnamefont {Ercius}},\ }\bibfield  {title} {\bibinfo
  {title} {Mapping electron beam-induced radiolytic damage in molecular
  crystals},\ }\href {https://doi.org/10.1093/mam/ozae044.902} {\bibfield
  {journal} {\bibinfo  {journal} {Microscopy and Microanalysis}\ }\textbf
  {\bibinfo {volume} {30}},\ \bibinfo {pages} {ozae044.902} (\bibinfo {year}
  {2024})},\ \Eprint
  {https://arxiv.org/abs/https://academic.oup.com/mam/article-pdf/30/Supplement\_1/ozae044.902/58669529/ozae044.902.pdf}
  {https://academic.oup.com/mam/article-pdf/30/Supplement\_1/ozae044.902/58669529/ozae044.902.pdf}
  \BibitemShut {NoStop}%
\bibitem [{\citenamefont {Kraft}(2000)}]{KRAFT20001}%
  \BibitemOpen
  \bibfield  {author} {\bibinfo {author} {\bibfnamefont {G.}~\bibnamefont
  {Kraft}},\ }\bibfield  {title} {\bibinfo {title} {Tumortherapy with ion
  beams},\ }\href
  {https://doi.org/https://doi.org/10.1016/S0168-9002(00)00802-0} {\bibfield
  {journal} {\bibinfo  {journal} {Nuclear Instruments and Methods in Physics
  Research Section A: Accelerators, Spectrometers, Detectors and Associated
  Equipment}\ }\textbf {\bibinfo {volume} {454}},\ \bibinfo {pages} {1}
  (\bibinfo {year} {2000})},\ \bibinfo {note} {proc. of the 1st Int Symp. on
  Applications of Particle Detectors in Medicine, Biology and
  Astrophysics}\BibitemShut {NoStop}%
\bibitem [{\citenamefont {Tommasino}\ and\ \citenamefont
  {Durante}(2015)}]{cancers7010353}%
  \BibitemOpen
  \bibfield  {author} {\bibinfo {author} {\bibfnamefont {F.}~\bibnamefont
  {Tommasino}}\ and\ \bibinfo {author} {\bibfnamefont {M.}~\bibnamefont
  {Durante}},\ }\bibfield  {title} {\bibinfo {title} {Proton radiobiology},\
  }\href {https://doi.org/10.3390/cancers7010353} {\bibfield  {journal}
  {\bibinfo  {journal} {Cancers}\ }\textbf {\bibinfo {volume} {7}},\ \bibinfo
  {pages} {353} (\bibinfo {year} {2015})}\BibitemShut {NoStop}%
\bibitem [{\citenamefont {Li}\ \emph {et~al.}(2024)\citenamefont {Li},
  \citenamefont {Li}, \citenamefont {Tian}, \citenamefont {Wang}, \citenamefont
  {Lin}, \citenamefont {Bai}, \citenamefont {Wang},\ and\ \citenamefont
  {Dong}}]{Li2024}%
  \BibitemOpen
  \bibfield  {author} {\bibinfo {author} {\bibfnamefont {Z.}~\bibnamefont
  {Li}}, \bibinfo {author} {\bibfnamefont {Q.}~\bibnamefont {Li}}, \bibinfo
  {author} {\bibfnamefont {H.}~\bibnamefont {Tian}}, \bibinfo {author}
  {\bibfnamefont {M.}~\bibnamefont {Wang}}, \bibinfo {author} {\bibfnamefont
  {R.}~\bibnamefont {Lin}}, \bibinfo {author} {\bibfnamefont {J.}~\bibnamefont
  {Bai}}, \bibinfo {author} {\bibfnamefont {D.}~\bibnamefont {Wang}},\ and\
  \bibinfo {author} {\bibfnamefont {M.}~\bibnamefont {Dong}},\ }\bibfield
  {title} {\bibinfo {title} {Proton beam therapy for craniopharyngioma: a
  systematic review and meta-analysis},\ }\href
  {https://doi.org/10.1186/s13014-024-02556-w} {\bibfield  {journal} {\bibinfo
  {journal} {Radiation Oncology}\ }\textbf {\bibinfo {volume} {19}},\ \bibinfo
  {pages} {161} (\bibinfo {year} {2024})}\BibitemShut {NoStop}%
\bibitem [{\citenamefont {Graeff}\ \emph {et~al.}(2023)\citenamefont {Graeff},
  \citenamefont {Volz},\ and\ \citenamefont {Durante}}]{GRAEFF2023104046}%
  \BibitemOpen
  \bibfield  {author} {\bibinfo {author} {\bibfnamefont {C.}~\bibnamefont
  {Graeff}}, \bibinfo {author} {\bibfnamefont {L.}~\bibnamefont {Volz}},\ and\
  \bibinfo {author} {\bibfnamefont {M.}~\bibnamefont {Durante}},\ }\bibfield
  {title} {\bibinfo {title} {Emerging technologies for cancer therapy using
  accelerated particles},\ }\href
  {https://doi.org/https://doi.org/10.1016/j.ppnp.2023.104046} {\bibfield
  {journal} {\bibinfo  {journal} {Progress in Particle and Nuclear Physics}\
  }\textbf {\bibinfo {volume} {131}},\ \bibinfo {pages} {104046} (\bibinfo
  {year} {2023})}\BibitemShut {NoStop}%
\bibitem [{\citenamefont {Huang}\ \emph {et~al.}(2024)\citenamefont {Huang},
  \citenamefont {Wu}, \citenamefont {Cai}, \citenamefont {Wu}, \citenamefont
  {Li}, \citenamefont {Jiang}, \citenamefont {Qiao}, \citenamefont {Jiang},\
  and\ \citenamefont {Ren}}]{Huang2024}%
  \BibitemOpen
  \bibfield  {author} {\bibinfo {author} {\bibfnamefont {L.}~\bibnamefont
  {Huang}}, \bibinfo {author} {\bibfnamefont {H.}~\bibnamefont {Wu}}, \bibinfo
  {author} {\bibfnamefont {G.}~\bibnamefont {Cai}}, \bibinfo {author}
  {\bibfnamefont {S.}~\bibnamefont {Wu}}, \bibinfo {author} {\bibfnamefont
  {D.}~\bibnamefont {Li}}, \bibinfo {author} {\bibfnamefont {T.}~\bibnamefont
  {Jiang}}, \bibinfo {author} {\bibfnamefont {B.}~\bibnamefont {Qiao}},
  \bibinfo {author} {\bibfnamefont {C.}~\bibnamefont {Jiang}},\ and\ \bibinfo
  {author} {\bibfnamefont {F.}~\bibnamefont {Ren}},\ }\bibfield  {title}
  {\bibinfo {title} {Recent progress in the application of ion beam technology
  in the modification and fabrication of nanostructured energy materials},\
  }\href {https://doi.org/10.1021/acsnano.3c07896} {\bibfield  {journal}
  {\bibinfo  {journal} {ACS Nano}\ }\textbf {\bibinfo {volume} {18}},\ \bibinfo
  {pages} {2578} (\bibinfo {year} {2024})}\BibitemShut {NoStop}%
\bibitem [{\citenamefont {Shabi}\ \emph {et~al.}(2025)\citenamefont {Shabi},
  \citenamefont {Girshevitz}, \citenamefont {Primetzhofer}, \citenamefont
  {Kaveh},\ and\ \citenamefont {Shlimak}}]{SHABI2025105872}%
  \BibitemOpen
  \bibfield  {author} {\bibinfo {author} {\bibfnamefont {N.}~\bibnamefont
  {Shabi}}, \bibinfo {author} {\bibfnamefont {O.}~\bibnamefont {Girshevitz}},
  \bibinfo {author} {\bibfnamefont {D.}~\bibnamefont {Primetzhofer}}, \bibinfo
  {author} {\bibfnamefont {M.}~\bibnamefont {Kaveh}},\ and\ \bibinfo {author}
  {\bibfnamefont {I.}~\bibnamefont {Shlimak}},\ }\bibfield  {title} {\bibinfo
  {title} {Dominant impact of ion velocity on defect formation in suspended
  graphene},\ }\href
  {https://doi.org/https://doi.org/10.1016/j.surfin.2025.105872} {\bibfield
  {journal} {\bibinfo  {journal} {Surfaces and Interfaces}\ }\textbf {\bibinfo
  {volume} {58}},\ \bibinfo {pages} {105872} (\bibinfo {year}
  {2025})}\BibitemShut {NoStop}%
\bibitem [{\citenamefont {Liu}\ \emph {et~al.}(2025)\citenamefont {Liu},
  \citenamefont {Deng}, \citenamefont {Wang}, \citenamefont {Wang},
  \citenamefont {Liu}, \citenamefont {Gong}, \citenamefont {Fan}, \citenamefont
  {Wei}, \citenamefont {Su}, \citenamefont {Wei}, \citenamefont {Wang},\ and\
  \citenamefont {Dan}}]{LIU2025163442}%
  \BibitemOpen
  \bibfield  {author} {\bibinfo {author} {\bibfnamefont {Y.}~\bibnamefont
  {Liu}}, \bibinfo {author} {\bibfnamefont {Y.}~\bibnamefont {Deng}}, \bibinfo
  {author} {\bibfnamefont {Y.}~\bibnamefont {Wang}}, \bibinfo {author}
  {\bibfnamefont {L.}~\bibnamefont {Wang}}, \bibinfo {author} {\bibfnamefont
  {T.}~\bibnamefont {Liu}}, \bibinfo {author} {\bibfnamefont {Z.}~\bibnamefont
  {Gong}}, \bibinfo {author} {\bibfnamefont {Z.}~\bibnamefont {Fan}}, \bibinfo
  {author} {\bibfnamefont {H.}~\bibnamefont {Wei}}, \bibinfo {author}
  {\bibfnamefont {Z.}~\bibnamefont {Su}}, \bibinfo {author} {\bibfnamefont
  {W.}~\bibnamefont {Wei}}, \bibinfo {author} {\bibfnamefont {Y.}~\bibnamefont
  {Wang}},\ and\ \bibinfo {author} {\bibfnamefont {Y.}~\bibnamefont {Dan}},\
  }\bibfield  {title} {\bibinfo {title} {Surface self-assembly of gold
  nanoparticles on graphite driven by ion-irradiation-induced atomic defects},\
  }\href {https://doi.org/https://doi.org/10.1016/j.apsusc.2025.163442}
  {\bibfield  {journal} {\bibinfo  {journal} {Applied Surface Science}\
  }\textbf {\bibinfo {volume} {704}},\ \bibinfo {pages} {163442} (\bibinfo
  {year} {2025})}\BibitemShut {NoStop}%
\bibitem [{\citenamefont {Krasheninnikov}\ and\ \citenamefont
  {Banhart}(2007)}]{Krasheninnikov2007}%
  \BibitemOpen
  \bibfield  {author} {\bibinfo {author} {\bibfnamefont {A.~V.}\ \bibnamefont
  {Krasheninnikov}}\ and\ \bibinfo {author} {\bibfnamefont {F.}~\bibnamefont
  {Banhart}},\ }\bibfield  {title} {\bibinfo {title} {Engineering of
  nanostructured carbon materials with electron or ion beams},\ }\href
  {https://doi.org/10.1038/nmat1996} {\bibfield  {journal} {\bibinfo  {journal}
  {Nature Materials}\ }\textbf {\bibinfo {volume} {6}},\ \bibinfo {pages} {723}
  (\bibinfo {year} {2007})}\BibitemShut {NoStop}%
\bibitem [{\citenamefont {Bubin}\ \emph {et~al.}(2012)\citenamefont {Bubin},
  \citenamefont {Wang}, \citenamefont {Pantelides},\ and\ \citenamefont
  {Varga}}]{PhysRevB.85.235435}%
  \BibitemOpen
  \bibfield  {author} {\bibinfo {author} {\bibfnamefont {S.}~\bibnamefont
  {Bubin}}, \bibinfo {author} {\bibfnamefont {B.}~\bibnamefont {Wang}},
  \bibinfo {author} {\bibfnamefont {S.}~\bibnamefont {Pantelides}},\ and\
  \bibinfo {author} {\bibfnamefont {K.}~\bibnamefont {Varga}},\ }\bibfield
  {title} {\bibinfo {title} {Simulation of high-energy ion collisions with
  graphene fragments},\ }\href {https://doi.org/10.1103/PhysRevB.85.235435}
  {\bibfield  {journal} {\bibinfo  {journal} {Phys. Rev. B}\ }\textbf {\bibinfo
  {volume} {85}},\ \bibinfo {pages} {235435} (\bibinfo {year}
  {2012})}\BibitemShut {NoStop}%
\bibitem [{\citenamefont {Wang}\ \emph {et~al.}(2015)\citenamefont {Wang},
  \citenamefont {Li},\ and\ \citenamefont {Wang}}]{PhysRevLett.114.063004}%
  \BibitemOpen
  \bibfield  {author} {\bibinfo {author} {\bibfnamefont {Z.}~\bibnamefont
  {Wang}}, \bibinfo {author} {\bibfnamefont {S.-S.}\ \bibnamefont {Li}},\ and\
  \bibinfo {author} {\bibfnamefont {L.-W.}\ \bibnamefont {Wang}},\ }\bibfield
  {title} {\bibinfo {title} {Efficient real-time time-dependent density
  functional theory method and its application to a collision of an ion with a
  2d material},\ }\href {https://doi.org/10.1103/PhysRevLett.114.063004}
  {\bibfield  {journal} {\bibinfo  {journal} {Phys. Rev. Lett.}\ }\textbf
  {\bibinfo {volume} {114}},\ \bibinfo {pages} {063004} (\bibinfo {year}
  {2015})}\BibitemShut {NoStop}%
\bibitem [{\citenamefont {Westlake}\ \emph {et~al.}(2014)\citenamefont
  {Westlake}, \citenamefont {Waite~Jr.}, \citenamefont {Carrasco},
  \citenamefont {Richard},\ and\ \citenamefont
  {Cravens}}]{https://doi.org/10.1002/2014JA020208}%
  \BibitemOpen
  \bibfield  {author} {\bibinfo {author} {\bibfnamefont {J.~H.}\ \bibnamefont
  {Westlake}}, \bibinfo {author} {\bibfnamefont {J.~H.}\ \bibnamefont
  {Waite~Jr.}}, \bibinfo {author} {\bibfnamefont {N.}~\bibnamefont {Carrasco}},
  \bibinfo {author} {\bibfnamefont {M.}~\bibnamefont {Richard}},\ and\ \bibinfo
  {author} {\bibfnamefont {T.}~\bibnamefont {Cravens}},\ }\bibfield  {title}
  {\bibinfo {title} {The role of ion-molecule reactions in the growth of heavy
  ions in titan's ionosphere},\ }\href
  {https://doi.org/https://doi.org/10.1002/2014JA020208} {\bibfield  {journal}
  {\bibinfo  {journal} {Journal of Geophysical Research: Space Physics}\
  }\textbf {\bibinfo {volume} {119}},\ \bibinfo {pages} {5951} (\bibinfo {year}
  {2014})},\ \Eprint
  {https://arxiv.org/abs/https://agupubs.onlinelibrary.wiley.com/doi/pdf/10.1002/2014JA020208}
  {https://agupubs.onlinelibrary.wiley.com/doi/pdf/10.1002/2014JA020208}
  \BibitemShut {NoStop}%
\bibitem [{\citenamefont {{Mancini, Luca}}\ \emph {et~al.}(2024)\citenamefont
  {{Mancini, Luca}}, \citenamefont {{Valença Ferreira de Aragao, Emilia}},
  \citenamefont {{Pirani, Fernando}}, \citenamefont {{Rosi, Marzio}},
  \citenamefont {{Faginas-Lago, Noelia}}, \citenamefont {{Richardson,
  Vincent}}, \citenamefont {{Martini, Luca Matteo}}, \citenamefont {{Podio,
  Linda}}, \citenamefont {{Lippi, Manuela}}, \citenamefont {{Codella,
  Claudio}},\ and\ \citenamefont {{Ascenzi, Daniela}}}]{refId0}%
  \BibitemOpen
  \bibfield  {author} {\bibinfo {author} {\bibnamefont {{Mancini, Luca}}},
  \bibinfo {author} {\bibnamefont {{Valença Ferreira de Aragao, Emilia}}},
  \bibinfo {author} {\bibnamefont {{Pirani, Fernando}}}, \bibinfo {author}
  {\bibnamefont {{Rosi, Marzio}}}, \bibinfo {author} {\bibnamefont
  {{Faginas-Lago, Noelia}}}, \bibinfo {author} {\bibnamefont {{Richardson,
  Vincent}}}, \bibinfo {author} {\bibnamefont {{Martini, Luca Matteo}}},
  \bibinfo {author} {\bibnamefont {{Podio, Linda}}}, \bibinfo {author}
  {\bibnamefont {{Lippi, Manuela}}}, \bibinfo {author} {\bibnamefont {{Codella,
  Claudio}}},\ and\ \bibinfo {author} {\bibnamefont {{Ascenzi, Daniela}}},\
  }\bibfield  {title} {\bibinfo {title} {Destruction of interstellar methyl
  cyanide (ch3cn) via collisions with he+ ions},\ }\href
  {https://doi.org/10.1051/0004-6361/202451674} {\bibfield  {journal} {\bibinfo
   {journal} {Astrophysics and Astronomy}\ }\textbf {\bibinfo {volume} {691}},\
  \bibinfo {pages} {A83} (\bibinfo {year} {2024})}\BibitemShut {NoStop}%
\bibitem [{\citenamefont {Cui}\ and\ \citenamefont {Herbst}(2024)}]{Cui2024}%
  \BibitemOpen
  \bibfield  {author} {\bibinfo {author} {\bibfnamefont {W.}~\bibnamefont
  {Cui}}\ and\ \bibinfo {author} {\bibfnamefont {E.}~\bibnamefont {Herbst}},\
  }\bibfield  {title} {\bibinfo {title} {Exploring the role of ion--molecule
  reactions on interstellar icy grain surfaces},\ }\href
  {https://doi.org/10.1021/acsearthspacechem.4c00194} {\bibfield  {journal}
  {\bibinfo  {journal} {ACS Earth and Space Chemistry}\ }\textbf {\bibinfo
  {volume} {8}},\ \bibinfo {pages} {2218} (\bibinfo {year} {2024})}\BibitemShut
  {NoStop}%
\bibitem [{\citenamefont {Lüdde}\ \emph {et~al.}(2019)\citenamefont {Lüdde},
  \citenamefont {Horbatsch},\ and\ \citenamefont {Kirchner}}]{Lüdde_2019}%
  \BibitemOpen
  \bibfield  {author} {\bibinfo {author} {\bibfnamefont {H.~J.}\ \bibnamefont
  {Lüdde}}, \bibinfo {author} {\bibfnamefont {M.}~\bibnamefont {Horbatsch}},\
  and\ \bibinfo {author} {\bibfnamefont {T.}~\bibnamefont {Kirchner}},\
  }\bibfield  {title} {\bibinfo {title} {Proton-impact-induced electron
  emission from biologically relevant molecules studied with a screened
  independent atom model},\ }\href {https://doi.org/10.1088/1361-6455/ab3a63}
  {\bibfield  {journal} {\bibinfo  {journal} {Journal of Physics B: Atomic,
  Molecular and Optical Physics}\ }\textbf {\bibinfo {volume} {52}},\ \bibinfo
  {pages} {195203} (\bibinfo {year} {2019})}\BibitemShut {NoStop}%
\bibitem [{\citenamefont {Luna}\ \emph {et~al.}(2007)\citenamefont {Luna},
  \citenamefont {de~Barros}, \citenamefont {Wyer}, \citenamefont {Scully},
  \citenamefont {Lecointre}, \citenamefont {Garcia}, \citenamefont {Sigaud},
  \citenamefont {Santos}, \citenamefont {Senthil}, \citenamefont {Shah},
  \citenamefont {Latimer},\ and\ \citenamefont
  {Montenegro}}]{PhysRevA.75.042711}%
  \BibitemOpen
  \bibfield  {author} {\bibinfo {author} {\bibfnamefont {H.}~\bibnamefont
  {Luna}}, \bibinfo {author} {\bibfnamefont {A.~L.~F.}\ \bibnamefont
  {de~Barros}}, \bibinfo {author} {\bibfnamefont {J.~A.}\ \bibnamefont {Wyer}},
  \bibinfo {author} {\bibfnamefont {S.~W.~J.}\ \bibnamefont {Scully}}, \bibinfo
  {author} {\bibfnamefont {J.}~\bibnamefont {Lecointre}}, \bibinfo {author}
  {\bibfnamefont {P.~M.~Y.}\ \bibnamefont {Garcia}}, \bibinfo {author}
  {\bibfnamefont {G.~M.}\ \bibnamefont {Sigaud}}, \bibinfo {author}
  {\bibfnamefont {A.~C.~F.}\ \bibnamefont {Santos}}, \bibinfo {author}
  {\bibfnamefont {V.}~\bibnamefont {Senthil}}, \bibinfo {author} {\bibfnamefont
  {M.~B.}\ \bibnamefont {Shah}}, \bibinfo {author} {\bibfnamefont {C.~J.}\
  \bibnamefont {Latimer}},\ and\ \bibinfo {author} {\bibfnamefont {E.~C.}\
  \bibnamefont {Montenegro}},\ }\bibfield  {title} {\bibinfo {title}
  {Water-molecule dissociation by proton and hydrogen impact},\ }\href
  {https://doi.org/10.1103/PhysRevA.75.042711} {\bibfield  {journal} {\bibinfo
  {journal} {Phys. Rev. A}\ }\textbf {\bibinfo {volume} {75}},\ \bibinfo
  {pages} {042711} (\bibinfo {year} {2007})}\BibitemShut {NoStop}%
\bibitem [{\citenamefont {Guan}\ and\ \citenamefont
  {Bartschat}(2009)}]{PhysRevLett.103.213201}%
  \BibitemOpen
  \bibfield  {author} {\bibinfo {author} {\bibfnamefont {X.}~\bibnamefont
  {Guan}}\ and\ \bibinfo {author} {\bibfnamefont {K.}~\bibnamefont
  {Bartschat}},\ }\bibfield  {title} {\bibinfo {title} {Complete breakup of the
  helium atom by proton and antiproton impact},\ }\href
  {https://doi.org/10.1103/PhysRevLett.103.213201} {\bibfield  {journal}
  {\bibinfo  {journal} {Phys. Rev. Lett.}\ }\textbf {\bibinfo {volume} {103}},\
  \bibinfo {pages} {213201} (\bibinfo {year} {2009})}\BibitemShut {NoStop}%
\bibitem [{\citenamefont {Abdurakhmanov}\ \emph {et~al.}(2018)\citenamefont
  {Abdurakhmanov}, \citenamefont {Alladustov}, \citenamefont {Bailey},
  \citenamefont {Kadyrov},\ and\ \citenamefont {Bray}}]{Abdurakhmanov_2018}%
  \BibitemOpen
  \bibfield  {author} {\bibinfo {author} {\bibfnamefont {I.~B.}\ \bibnamefont
  {Abdurakhmanov}}, \bibinfo {author} {\bibfnamefont {S.~U.}\ \bibnamefont
  {Alladustov}}, \bibinfo {author} {\bibfnamefont {J.~J.}\ \bibnamefont
  {Bailey}}, \bibinfo {author} {\bibfnamefont {A.~S.}\ \bibnamefont
  {Kadyrov}},\ and\ \bibinfo {author} {\bibfnamefont {I.}~\bibnamefont
  {Bray}},\ }\bibfield  {title} {\bibinfo {title} {Proton scattering from
  excited states of atomic hydrogen},\ }\href
  {https://doi.org/10.1088/1361-6587/aad436} {\bibfield  {journal} {\bibinfo
  {journal} {Plasma Physics and Controlled Fusion}\ }\textbf {\bibinfo {volume}
  {60}},\ \bibinfo {pages} {095009} (\bibinfo {year} {2018})}\BibitemShut
  {NoStop}%
\bibitem [{\citenamefont {Leung}\ and\ \citenamefont
  {Kirchner}(2019)}]{Leung2019}%
  \BibitemOpen
  \bibfield  {author} {\bibinfo {author} {\bibfnamefont {A.~C.~K.}\
  \bibnamefont {Leung}}\ and\ \bibinfo {author} {\bibfnamefont
  {T.}~\bibnamefont {Kirchner}},\ }\bibfield  {title} {\bibinfo {title} {Proton
  impact on ground and excited states of atomic hydrogen},\ }\href
  {https://doi.org/10.1140/epjd/e2019-100380-x} {\bibfield  {journal} {\bibinfo
   {journal} {The European Physical Journal D}\ }\textbf {\bibinfo {volume}
  {73}},\ \bibinfo {pages} {246} (\bibinfo {year} {2019})}\BibitemShut
  {NoStop}%
\bibitem [{\citenamefont {Seraide}\ \emph {et~al.}(2017)\citenamefont
  {Seraide}, \citenamefont {Bernal}, \citenamefont {Brunetto}, \citenamefont
  {de~Giovannini},\ and\ \citenamefont {Rubio}}]{Seraide2017}%
  \BibitemOpen
  \bibfield  {author} {\bibinfo {author} {\bibfnamefont {R.}~\bibnamefont
  {Seraide}}, \bibinfo {author} {\bibfnamefont {M.~A.}\ \bibnamefont {Bernal}},
  \bibinfo {author} {\bibfnamefont {G.}~\bibnamefont {Brunetto}}, \bibinfo
  {author} {\bibfnamefont {U.}~\bibnamefont {de~Giovannini}},\ and\ \bibinfo
  {author} {\bibfnamefont {A.}~\bibnamefont {Rubio}},\ }\bibfield  {title}
  {\bibinfo {title} {Tddft-based study on the proton--dna collision},\ }\href
  {https://doi.org/10.1021/acs.jpcb.7b04934} {\bibfield  {journal} {\bibinfo
  {journal} {The Journal of Physical Chemistry B}\ }\textbf {\bibinfo {volume}
  {121}},\ \bibinfo {pages} {7276} (\bibinfo {year} {2017})}\BibitemShut
  {NoStop}%
\bibitem [{\citenamefont {Covington}\ \emph {et~al.}(2017)\citenamefont
  {Covington}, \citenamefont {Hartig}, \citenamefont {Russakoff}, \citenamefont
  {Kulpins},\ and\ \citenamefont {Varga}}]{PhysRevA.95.052701}%
  \BibitemOpen
  \bibfield  {author} {\bibinfo {author} {\bibfnamefont {C.}~\bibnamefont
  {Covington}}, \bibinfo {author} {\bibfnamefont {K.}~\bibnamefont {Hartig}},
  \bibinfo {author} {\bibfnamefont {A.}~\bibnamefont {Russakoff}}, \bibinfo
  {author} {\bibfnamefont {R.}~\bibnamefont {Kulpins}},\ and\ \bibinfo {author}
  {\bibfnamefont {K.}~\bibnamefont {Varga}},\ }\bibfield  {title} {\bibinfo
  {title} {Time-dependent density-functional-theory investigation of the
  collisions of protons and $\ensuremath{\alpha}$ particles with uracil and
  adenine},\ }\href {https://doi.org/10.1103/PhysRevA.95.052701} {\bibfield
  {journal} {\bibinfo  {journal} {Phys. Rev. A}\ }\textbf {\bibinfo {volume}
  {95}},\ \bibinfo {pages} {052701} (\bibinfo {year} {2017})}\BibitemShut
  {NoStop}%
\bibitem [{\citenamefont {Wang}\ \emph {et~al.}(2022)\citenamefont {Wang},
  \citenamefont {Wang}, \citenamefont {Zhang},\ and\ \citenamefont
  {Qian}}]{Wang_2022}%
  \BibitemOpen
  \bibfield  {author} {\bibinfo {author} {\bibfnamefont {X.}~\bibnamefont
  {Wang}}, \bibinfo {author} {\bibfnamefont {Z.-P.}\ \bibnamefont {Wang}},
  \bibinfo {author} {\bibfnamefont {F.-S.}\ \bibnamefont {Zhang}},\ and\
  \bibinfo {author} {\bibfnamefont {C.-Y.}\ \bibnamefont {Qian}},\ }\bibfield
  {title} {\bibinfo {title} {Collision site effect on the radiation dynamics of
  cytosine induced by proton},\ }\href
  {https://doi.org/10.1088/1674-1056/ac4900} {\bibfield  {journal} {\bibinfo
  {journal} {Chinese Physics B}\ }\textbf {\bibinfo {volume} {31}},\ \bibinfo
  {pages} {063401} (\bibinfo {year} {2022})}\BibitemShut {NoStop}%
\bibitem [{\citenamefont {Sanders}\ and\ \citenamefont
  {Varghese}(1999)}]{10.1063/1.59216}%
  \BibitemOpen
  \bibfield  {author} {\bibinfo {author} {\bibfnamefont {J.~M.}\ \bibnamefont
  {Sanders}}\ and\ \bibinfo {author} {\bibfnamefont {S.~L.}\ \bibnamefont
  {Varghese}},\ }\bibfield  {title} {\bibinfo {title} {Electron capture from
  hydrocarbon molecules by proton projectiles in the 60–120 kev energy
  range},\ }\href {https://doi.org/10.1063/1.59216} {\bibfield  {journal}
  {\bibinfo  {journal} {AIP Conference Proceedings}\ }\textbf {\bibinfo
  {volume} {475}},\ \bibinfo {pages} {70} (\bibinfo {year} {1999})},\ \Eprint
  {https://arxiv.org/abs/https://pubs.aip.org/aip/acp/article-pdf/475/1/70/12099216/70\_1\_online.pdf}
  {https://pubs.aip.org/aip/acp/article-pdf/475/1/70/12099216/70\_1\_online.pdf}
  \BibitemShut {NoStop}%
\bibitem [{\citenamefont {Sanders}\ \emph {et~al.}(2003)\citenamefont
  {Sanders}, \citenamefont {Varghese}, \citenamefont {Fleming},\ and\
  \citenamefont {Soosai}}]{J_M_Sanders_2003}%
  \BibitemOpen
  \bibfield  {author} {\bibinfo {author} {\bibfnamefont {J.~M.}\ \bibnamefont
  {Sanders}}, \bibinfo {author} {\bibfnamefont {S.~L.}\ \bibnamefont
  {Varghese}}, \bibinfo {author} {\bibfnamefont {C.~H.}\ \bibnamefont
  {Fleming}},\ and\ \bibinfo {author} {\bibfnamefont {G.~A.}\ \bibnamefont
  {Soosai}},\ }\bibfield  {title} {\bibinfo {title} {Electron capture by
  protons and electron loss from hydrogen atoms in collisions with hydrocarbon
  and hydrogen molecules in the 60–120 kev energy range},\ }\href
  {https://doi.org/10.1088/0953-4075/36/18/311} {\bibfield  {journal} {\bibinfo
   {journal} {Journal of Physics B: Atomic, Molecular and Optical Physics}\
  }\textbf {\bibinfo {volume} {36}},\ \bibinfo {pages} {3835} (\bibinfo {year}
  {2003})}\BibitemShut {NoStop}%
\bibitem [{\citenamefont {Baumgart}\ \emph {et~al.}(1984)\citenamefont
  {Baumgart}, \citenamefont {Arnold}, \citenamefont {Günzl}, \citenamefont
  {Huttel}, \citenamefont {Hofmann}, \citenamefont {Kniest}, \citenamefont
  {Pfaff}, \citenamefont {Reiter}, \citenamefont {Tharraketta},\ and\
  \citenamefont {Clausnitzer}}]{BAUMGART19841}%
  \BibitemOpen
  \bibfield  {author} {\bibinfo {author} {\bibfnamefont {H.}~\bibnamefont
  {Baumgart}}, \bibinfo {author} {\bibfnamefont {W.}~\bibnamefont {Arnold}},
  \bibinfo {author} {\bibfnamefont {J.}~\bibnamefont {Günzl}}, \bibinfo
  {author} {\bibfnamefont {E.}~\bibnamefont {Huttel}}, \bibinfo {author}
  {\bibfnamefont {A.}~\bibnamefont {Hofmann}}, \bibinfo {author} {\bibfnamefont
  {N.}~\bibnamefont {Kniest}}, \bibinfo {author} {\bibfnamefont
  {E.}~\bibnamefont {Pfaff}}, \bibinfo {author} {\bibfnamefont
  {G.}~\bibnamefont {Reiter}}, \bibinfo {author} {\bibfnamefont
  {S.}~\bibnamefont {Tharraketta}},\ and\ \bibinfo {author} {\bibfnamefont
  {G.}~\bibnamefont {Clausnitzer}},\ }\bibfield  {title} {\bibinfo {title}
  {Proton and helium stopping cross sections in gaseous hydrocarbon
  compounds},\ }\href
  {https://doi.org/https://doi.org/10.1016/0168-583X(84)90561-5} {\bibfield
  {journal} {\bibinfo  {journal} {Nuclear Instruments and Methods in Physics
  Research Section B: Beam Interactions with Materials and Atoms}\ }\textbf
  {\bibinfo {volume} {5}},\ \bibinfo {pages} {1} (\bibinfo {year}
  {1984})}\BibitemShut {NoStop}%
\bibitem [{\citenamefont {Lüdde}\ \emph {et~al.}(2020)\citenamefont {Lüdde},
  \citenamefont {Jorge}, \citenamefont {Horbatsch},\ and\ \citenamefont
  {Kirchner}}]{atoms8030059}%
  \BibitemOpen
  \bibfield  {author} {\bibinfo {author} {\bibfnamefont {H.~J.}\ \bibnamefont
  {Lüdde}}, \bibinfo {author} {\bibfnamefont {A.}~\bibnamefont {Jorge}},
  \bibinfo {author} {\bibfnamefont {M.}~\bibnamefont {Horbatsch}},\ and\
  \bibinfo {author} {\bibfnamefont {T.}~\bibnamefont {Kirchner}},\ }\bibfield
  {title} {\bibinfo {title} {Net electron capture in collisions of multiply
  charged projectiles with biologically relevant molecules},\ }\bibfield
  {journal} {\bibinfo  {journal} {Atoms}\ }\textbf {\bibinfo {volume} {8}},\
  \href {https://doi.org/10.3390/atoms8030059} {10.3390/atoms8030059} (\bibinfo
  {year} {2020})\BibitemShut {NoStop}%
\bibitem [{\citenamefont {L{\"u}dde}\ \emph {et~al.}(2019)\citenamefont
  {L{\"u}dde}, \citenamefont {Horbatsch},\ and\ \citenamefont
  {Kirchner}}]{Lüdde2019}%
  \BibitemOpen
  \bibfield  {author} {\bibinfo {author} {\bibfnamefont {H.~J.}\ \bibnamefont
  {L{\"u}dde}}, \bibinfo {author} {\bibfnamefont {M.}~\bibnamefont
  {Horbatsch}},\ and\ \bibinfo {author} {\bibfnamefont {T.}~\bibnamefont
  {Kirchner}},\ }\bibfield  {title} {\bibinfo {title} {Electron capture and
  ionization cross-section calculations for proton collisions with methane and
  the dna and rna nucleobases},\ }\href
  {https://doi.org/10.1140/epjd/e2019-100344-2} {\bibfield  {journal} {\bibinfo
   {journal} {The European Physical Journal D}\ }\textbf {\bibinfo {volume}
  {73}},\ \bibinfo {pages} {249} (\bibinfo {year} {2019})}\BibitemShut
  {NoStop}%
\bibitem [{\citenamefont {Roither}\ \emph {et~al.}(2011)\citenamefont
  {Roither}, \citenamefont {Xie}, \citenamefont {Kartashov}, \citenamefont
  {Zhang}, \citenamefont {Sch\"offler}, \citenamefont {Xu}, \citenamefont
  {Iwasaki}, \citenamefont {Okino}, \citenamefont {Yamanouchi}, \citenamefont
  {Baltuska},\ and\ \citenamefont {Kitzler}}]{PhysRevLett.106.163001}%
  \BibitemOpen
  \bibfield  {author} {\bibinfo {author} {\bibfnamefont {S.}~\bibnamefont
  {Roither}}, \bibinfo {author} {\bibfnamefont {X.}~\bibnamefont {Xie}},
  \bibinfo {author} {\bibfnamefont {D.}~\bibnamefont {Kartashov}}, \bibinfo
  {author} {\bibfnamefont {L.}~\bibnamefont {Zhang}}, \bibinfo {author}
  {\bibfnamefont {M.}~\bibnamefont {Sch\"offler}}, \bibinfo {author}
  {\bibfnamefont {H.}~\bibnamefont {Xu}}, \bibinfo {author} {\bibfnamefont
  {A.}~\bibnamefont {Iwasaki}}, \bibinfo {author} {\bibfnamefont
  {T.}~\bibnamefont {Okino}}, \bibinfo {author} {\bibfnamefont
  {K.}~\bibnamefont {Yamanouchi}}, \bibinfo {author} {\bibfnamefont
  {A.}~\bibnamefont {Baltuska}},\ and\ \bibinfo {author} {\bibfnamefont
  {M.}~\bibnamefont {Kitzler}},\ }\bibfield  {title} {\bibinfo {title} {High
  energy proton ejection from hydrocarbon molecules driven by highly efficient
  field ionization},\ }\href {https://doi.org/10.1103/PhysRevLett.106.163001}
  {\bibfield  {journal} {\bibinfo  {journal} {Phys. Rev. Lett.}\ }\textbf
  {\bibinfo {volume} {106}},\ \bibinfo {pages} {163001} (\bibinfo {year}
  {2011})}\BibitemShut {NoStop}%
\bibitem [{\citenamefont {Zhang}\ \emph {et~al.}(2019)\citenamefont {Zhang},
  \citenamefont {Wang}, \citenamefont {Wei}, \citenamefont {Jiang},
  \citenamefont {Yu}, \citenamefont {Hutton}, \citenamefont {Zou},
  \citenamefont {Chen},\ and\ \citenamefont {Wei}}]{10.1063/1.5088690}%
  \BibitemOpen
  \bibfield  {author} {\bibinfo {author} {\bibfnamefont {Y.}~\bibnamefont
  {Zhang}}, \bibinfo {author} {\bibfnamefont {B.}~\bibnamefont {Wang}},
  \bibinfo {author} {\bibfnamefont {L.}~\bibnamefont {Wei}}, \bibinfo {author}
  {\bibfnamefont {T.}~\bibnamefont {Jiang}}, \bibinfo {author} {\bibfnamefont
  {W.}~\bibnamefont {Yu}}, \bibinfo {author} {\bibfnamefont {R.}~\bibnamefont
  {Hutton}}, \bibinfo {author} {\bibfnamefont {Y.}~\bibnamefont {Zou}},
  \bibinfo {author} {\bibfnamefont {L.}~\bibnamefont {Chen}},\ and\ \bibinfo
  {author} {\bibfnamefont {B.}~\bibnamefont {Wei}},\ }\bibfield  {title}
  {\bibinfo {title} {Proton migration in hydrocarbons induced by slow highly
  charged ion impact},\ }\href {https://doi.org/10.1063/1.5088690} {\bibfield
  {journal} {\bibinfo  {journal} {The Journal of Chemical Physics}\ }\textbf
  {\bibinfo {volume} {150}},\ \bibinfo {pages} {204303} (\bibinfo {year}
  {2019})},\ \Eprint
  {https://arxiv.org/abs/https://pubs.aip.org/aip/jcp/article-pdf/doi/10.1063/1.5088690/9651872/204303\_1\_online.pdf}
  {https://pubs.aip.org/aip/jcp/article-pdf/doi/10.1063/1.5088690/9651872/204303\_1\_online.pdf}
  \BibitemShut {NoStop}%
\bibitem [{\citenamefont {Quashie}\ \emph {et~al.}(2017)\citenamefont
  {Quashie}, \citenamefont {Saha}, \citenamefont {Andrade},\ and\ \citenamefont
  {Correa}}]{PhysRevA.95.042517}%
  \BibitemOpen
  \bibfield  {author} {\bibinfo {author} {\bibfnamefont {E.~E.}\ \bibnamefont
  {Quashie}}, \bibinfo {author} {\bibfnamefont {B.~C.}\ \bibnamefont {Saha}},
  \bibinfo {author} {\bibfnamefont {X.}~\bibnamefont {Andrade}},\ and\ \bibinfo
  {author} {\bibfnamefont {A.~A.}\ \bibnamefont {Correa}},\ }\bibfield  {title}
  {\bibinfo {title} {Self-interaction effects on charge-transfer collisions},\
  }\href {https://doi.org/10.1103/PhysRevA.95.042517} {\bibfield  {journal}
  {\bibinfo  {journal} {Phys. Rev. A}\ }\textbf {\bibinfo {volume} {95}},\
  \bibinfo {pages} {042517} (\bibinfo {year} {2017})}\BibitemShut {NoStop}%
\bibitem [{\citenamefont {Herv{\'e}~du Penhoat}\ \emph
  {et~al.}(2018)\citenamefont {Herv{\'e}~du Penhoat}, \citenamefont {Moraga},
  \citenamefont {Gaigeot}, \citenamefont {Vuilleumier}, \citenamefont
  {Tavernelli},\ and\ \citenamefont {Politis}}]{herve_du_penhoat_proton_2018}%
  \BibitemOpen
  \bibfield  {author} {\bibinfo {author} {\bibfnamefont {M.-A.}\ \bibnamefont
  {Herv{\'e}~du Penhoat}}, \bibinfo {author} {\bibfnamefont {N.~R.}\
  \bibnamefont {Moraga}}, \bibinfo {author} {\bibfnamefont {M.-P.}\
  \bibnamefont {Gaigeot}}, \bibinfo {author} {\bibfnamefont {R.}~\bibnamefont
  {Vuilleumier}}, \bibinfo {author} {\bibfnamefont {I.}~\bibnamefont
  {Tavernelli}},\ and\ \bibinfo {author} {\bibfnamefont {M.-F.}\ \bibnamefont
  {Politis}},\ }\bibfield  {title} {\bibinfo {title} {Proton {Collision} on
  {Deoxyribose} {Originating} from {Doubly} {Ionized} {Water} {Molecule}
  {Dissociation}},\ }\href {https://doi.org/10.1021/acs.jpca.8b04787}
  {\bibfield  {journal} {\bibinfo  {journal} {The Journal of Physical Chemistry
  A}\ }\textbf {\bibinfo {volume} {122}},\ \bibinfo {pages} {5311} (\bibinfo
  {year} {2018})},\ \bibinfo {note} {publisher: American Chemical
  Society}\BibitemShut {NoStop}%
\bibitem [{\citenamefont {Miyamoto}\ and\ \citenamefont
  {Komatsu}(2019)}]{miyamoto_molecular-scale_2019}%
  \BibitemOpen
  \bibfield  {author} {\bibinfo {author} {\bibfnamefont {Y.}~\bibnamefont
  {Miyamoto}}\ and\ \bibinfo {author} {\bibfnamefont {T.}~\bibnamefont
  {Komatsu}},\ }\bibfield  {title} {\bibinfo {title} {Molecular-scale modeling
  of light emission by combustion: An ab initio study},\ }\href
  {https://doi.org/10.1038/s41598-019-49200-2} {\bibfield  {journal} {\bibinfo
  {journal} {Scientific Reports}\ }\textbf {\bibinfo {volume} {9}},\ \bibinfo
  {pages} {12707} (\bibinfo {year} {2019})}\BibitemShut {NoStop}%
\bibitem [{\citenamefont {Gao}\ \emph {et~al.}(2014)\citenamefont {Gao},
  \citenamefont {Wang}, \citenamefont {Wang},\ and\ \citenamefont
  {Zhang}}]{gao_theoretical_2014}%
  \BibitemOpen
  \bibfield  {author} {\bibinfo {author} {\bibfnamefont {C.-Z.}\ \bibnamefont
  {Gao}}, \bibinfo {author} {\bibfnamefont {J.}~\bibnamefont {Wang}}, \bibinfo
  {author} {\bibfnamefont {F.}~\bibnamefont {Wang}},\ and\ \bibinfo {author}
  {\bibfnamefont {F.-S.}\ \bibnamefont {Zhang}},\ }\bibfield  {title} {\bibinfo
  {title} {Theoretical study on collision dynamics of {H}+ + {CH4} at low
  energies},\ }\href {https://doi.org/10.1063/1.4863635} {\bibfield  {journal}
  {\bibinfo  {journal} {The Journal of Chemical Physics}\ }\textbf {\bibinfo
  {volume} {140}},\ \bibinfo {pages} {054308} (\bibinfo {year}
  {2014})}\BibitemShut {NoStop}%
\bibitem [{\citenamefont {Taylor}\ \emph
  {et~al.}(2025{\natexlab{a}})\citenamefont {Taylor}, \citenamefont {Varga},
  \citenamefont {Mogyor\'osi}, \citenamefont {Chik\'an},\ and\ \citenamefont
  {Covington}}]{PhysRevA.111.013109}%
  \BibitemOpen
  \bibfield  {author} {\bibinfo {author} {\bibfnamefont {S.~S.}\ \bibnamefont
  {Taylor}}, \bibinfo {author} {\bibfnamefont {K.}~\bibnamefont {Varga}},
  \bibinfo {author} {\bibfnamefont {K.}~\bibnamefont {Mogyor\'osi}}, \bibinfo
  {author} {\bibfnamefont {V.}~\bibnamefont {Chik\'an}},\ and\ \bibinfo
  {author} {\bibfnamefont {C.}~\bibnamefont {Covington}},\ }\bibfield  {title}
  {\bibinfo {title} {Fragmentation in coulomb explosion of hydrocarbon
  molecules},\ }\href {https://doi.org/10.1103/PhysRevA.111.013109} {\bibfield
  {journal} {\bibinfo  {journal} {Phys. Rev. A}\ }\textbf {\bibinfo {volume}
  {111}},\ \bibinfo {pages} {013109} (\bibinfo {year}
  {2025}{\natexlab{a}})}\BibitemShut {NoStop}%
\bibitem [{\citenamefont {Last}\ and\ \citenamefont
  {Jortner}(2002)}]{Last2002}%
  \BibitemOpen
  \bibfield  {author} {\bibinfo {author} {\bibfnamefont {I.}~\bibnamefont
  {Last}}\ and\ \bibinfo {author} {\bibfnamefont {J.}~\bibnamefont {Jortner}},\
  }\bibfield  {title} {\bibinfo {title} {Nuclear fusion driven by coulomb
  explosion of methane clusters},\ }\href {https://doi.org/10.1021/jp0206121}
  {\bibfield  {journal} {\bibinfo  {journal} {The Journal of Physical Chemistry
  A}\ }\textbf {\bibinfo {volume} {106}},\ \bibinfo {pages} {10877} (\bibinfo
  {year} {2002})}\BibitemShut {NoStop}%
\bibitem [{\citenamefont {Cornaggia}\ \emph {et~al.}(1992)\citenamefont
  {Cornaggia}, \citenamefont {Normand},\ and\ \citenamefont
  {Morellec}}]{Cornaggia_1992}%
  \BibitemOpen
  \bibfield  {author} {\bibinfo {author} {\bibfnamefont {C.}~\bibnamefont
  {Cornaggia}}, \bibinfo {author} {\bibfnamefont {D.}~\bibnamefont {Normand}},\
  and\ \bibinfo {author} {\bibfnamefont {J.}~\bibnamefont {Morellec}},\
  }\bibfield  {title} {\bibinfo {title} {Role of the molecular electronic
  configuration in the coulomb fragmentation of n2, c2h2 and c2h4 in an intense
  laser field},\ }\href {https://doi.org/10.1088/0953-4075/25/17/003}
  {\bibfield  {journal} {\bibinfo  {journal} {Journal of Physics B: Atomic,
  Molecular and Optical Physics}\ }\textbf {\bibinfo {volume} {25}},\ \bibinfo
  {pages} {L415} (\bibinfo {year} {1992})}\BibitemShut {NoStop}%
\bibitem [{\citenamefont {Mogyor{\'o}si}\ \emph {et~al.}(2025)\citenamefont
  {Mogyor{\'o}si}, \citenamefont {T{\'o}th}, \citenamefont {S{\'a}rosi},
  \citenamefont {Gilicze}, \citenamefont {Csontos}, \citenamefont
  {Somosk{\H{o}}i}, \citenamefont {T{\'o}th}, \citenamefont {Prasannan~Geetha},
  \citenamefont {T{\'o}th}, \citenamefont {Taylor}, \citenamefont {Skoufis},
  \citenamefont {Barron}, \citenamefont {Varga}, \citenamefont {Covington},\
  and\ \citenamefont {Chik{\'a}n}}]{Mogyorosi2025}%
  \BibitemOpen
  \bibfield  {author} {\bibinfo {author} {\bibfnamefont {K.}~\bibnamefont
  {Mogyor{\'o}si}}, \bibinfo {author} {\bibfnamefont {B.}~\bibnamefont
  {T{\'o}th}}, \bibinfo {author} {\bibfnamefont {K.}~\bibnamefont
  {S{\'a}rosi}}, \bibinfo {author} {\bibfnamefont {B.}~\bibnamefont {Gilicze}},
  \bibinfo {author} {\bibfnamefont {J.}~\bibnamefont {Csontos}}, \bibinfo
  {author} {\bibfnamefont {T.}~\bibnamefont {Somosk{\H{o}}i}}, \bibinfo
  {author} {\bibfnamefont {S.}~\bibnamefont {T{\'o}th}}, \bibinfo {author}
  {\bibfnamefont {P.}~\bibnamefont {Prasannan~Geetha}}, \bibinfo {author}
  {\bibfnamefont {L.}~\bibnamefont {T{\'o}th}}, \bibinfo {author}
  {\bibfnamefont {S.~S.}\ \bibnamefont {Taylor}}, \bibinfo {author}
  {\bibfnamefont {N.}~\bibnamefont {Skoufis}}, \bibinfo {author} {\bibfnamefont
  {L.}~\bibnamefont {Barron}}, \bibinfo {author} {\bibfnamefont
  {K.}~\bibnamefont {Varga}}, \bibinfo {author} {\bibfnamefont
  {C.}~\bibnamefont {Covington}},\ and\ \bibinfo {author} {\bibfnamefont
  {V.}~\bibnamefont {Chik{\'a}n}},\ }\bibfield  {title} {\bibinfo {title}
  {Ch(a) radical formation in coulomb explosion from butane seeded plasma
  generated with chirp-controlled ultrashort laser pulses},\ }\href
  {https://doi.org/10.1021/acsomega.4c11074} {\bibfield  {journal} {\bibinfo
  {journal} {ACS Omega}\ }\textbf {\bibinfo {volume} {10}},\ \bibinfo {pages}
  {25285} (\bibinfo {year} {2025})}\BibitemShut {NoStop}%
\bibitem [{\citenamefont {Markevitch}\ \emph {et~al.}(2004)\citenamefont
  {Markevitch}, \citenamefont {Romanov}, \citenamefont {Smith},\ and\
  \citenamefont {Levis}}]{PhysRevLett.92.063001}%
  \BibitemOpen
  \bibfield  {author} {\bibinfo {author} {\bibfnamefont {A.~N.}\ \bibnamefont
  {Markevitch}}, \bibinfo {author} {\bibfnamefont {D.~A.}\ \bibnamefont
  {Romanov}}, \bibinfo {author} {\bibfnamefont {S.~M.}\ \bibnamefont {Smith}},\
  and\ \bibinfo {author} {\bibfnamefont {R.~J.}\ \bibnamefont {Levis}},\
  }\bibfield  {title} {\bibinfo {title} {Coulomb explosion of large polyatomic
  molecules assisted by nonadiabatic charge localization},\ }\href
  {https://doi.org/10.1103/PhysRevLett.92.063001} {\bibfield  {journal}
  {\bibinfo  {journal} {Phys. Rev. Lett.}\ }\textbf {\bibinfo {volume} {92}},\
  \bibinfo {pages} {063001} (\bibinfo {year} {2004})}\BibitemShut {NoStop}%
\bibitem [{\citenamefont {Hasegawa}\ \emph {et~al.}(2023)\citenamefont
  {Hasegawa}, \citenamefont {Matsuda}, \citenamefont {Morishita}, \citenamefont
  {Madsen}, \citenamefont {Jensen}, \citenamefont {Tolstikhin},\ and\
  \citenamefont {Hishikawa}}]{D3CP02337K}%
  \BibitemOpen
  \bibfield  {author} {\bibinfo {author} {\bibfnamefont {H.}~\bibnamefont
  {Hasegawa}}, \bibinfo {author} {\bibfnamefont {A.}~\bibnamefont {Matsuda}},
  \bibinfo {author} {\bibfnamefont {T.}~\bibnamefont {Morishita}}, \bibinfo
  {author} {\bibfnamefont {L.~B.}\ \bibnamefont {Madsen}}, \bibinfo {author}
  {\bibfnamefont {F.}~\bibnamefont {Jensen}}, \bibinfo {author} {\bibfnamefont
  {O.~I.}\ \bibnamefont {Tolstikhin}},\ and\ \bibinfo {author} {\bibfnamefont
  {A.}~\bibnamefont {Hishikawa}},\ }\bibfield  {title} {\bibinfo {title}
  {Dissociative ionization and coulomb explosion of ch4 in two-color asymmetric
  intense laser fields},\ }\href {https://doi.org/10.1039/D3CP02337K}
  {\bibfield  {journal} {\bibinfo  {journal} {Phys. Chem. Chem. Phys.}\
  }\textbf {\bibinfo {volume} {25}},\ \bibinfo {pages} {25408} (\bibinfo {year}
  {2023})}\BibitemShut {NoStop}%
\bibitem [{\citenamefont {Taylor}\ \emph
  {et~al.}(2025{\natexlab{b}})\citenamefont {Taylor}, \citenamefont
  {Covington},\ and\ \citenamefont {Varga}}]{PhysRevA.111.033109}%
  \BibitemOpen
  \bibfield  {author} {\bibinfo {author} {\bibfnamefont {S.~S.}\ \bibnamefont
  {Taylor}}, \bibinfo {author} {\bibfnamefont {C.}~\bibnamefont {Covington}},\
  and\ \bibinfo {author} {\bibfnamefont {K.}~\bibnamefont {Varga}},\ }\bibfield
   {title} {\bibinfo {title} {Quantum effects of coulomb explosion simulations
  revealed by time-dependent density-functional theory},\ }\href
  {https://doi.org/10.1103/PhysRevA.111.033109} {\bibfield  {journal} {\bibinfo
   {journal} {Phys. Rev. A}\ }\textbf {\bibinfo {volume} {111}},\ \bibinfo
  {pages} {033109} (\bibinfo {year} {2025}{\natexlab{b}})}\BibitemShut
  {NoStop}%
\bibitem [{\citenamefont {Palaniyappan}\ \emph {et~al.}(2010)\citenamefont
  {Palaniyappan}, \citenamefont {Mitchell}, \citenamefont {Ekanayake},
  \citenamefont {Watts}, \citenamefont {White}, \citenamefont {Sauer},
  \citenamefont {Howard}, \citenamefont {Videtto}, \citenamefont {Mancuso},
  \citenamefont {Wells}, \citenamefont {Stanev}, \citenamefont {Wen},
  \citenamefont {Decamp},\ and\ \citenamefont {Walker}}]{PhysRevA.82.043433}%
  \BibitemOpen
  \bibfield  {author} {\bibinfo {author} {\bibfnamefont {S.}~\bibnamefont
  {Palaniyappan}}, \bibinfo {author} {\bibfnamefont {R.}~\bibnamefont
  {Mitchell}}, \bibinfo {author} {\bibfnamefont {N.}~\bibnamefont {Ekanayake}},
  \bibinfo {author} {\bibfnamefont {A.~M.}\ \bibnamefont {Watts}}, \bibinfo
  {author} {\bibfnamefont {S.~L.}\ \bibnamefont {White}}, \bibinfo {author}
  {\bibfnamefont {R.}~\bibnamefont {Sauer}}, \bibinfo {author} {\bibfnamefont
  {L.~E.}\ \bibnamefont {Howard}}, \bibinfo {author} {\bibfnamefont
  {M.}~\bibnamefont {Videtto}}, \bibinfo {author} {\bibfnamefont
  {C.}~\bibnamefont {Mancuso}}, \bibinfo {author} {\bibfnamefont {S.~J.}\
  \bibnamefont {Wells}}, \bibinfo {author} {\bibfnamefont {T.}~\bibnamefont
  {Stanev}}, \bibinfo {author} {\bibfnamefont {B.~L.}\ \bibnamefont {Wen}},
  \bibinfo {author} {\bibfnamefont {M.~F.}\ \bibnamefont {Decamp}},\ and\
  \bibinfo {author} {\bibfnamefont {B.~C.}\ \bibnamefont {Walker}},\ }\bibfield
   {title} {\bibinfo {title} {Ionization of ethane, butane, and octane in
  strong laser fields},\ }\href {https://doi.org/10.1103/PhysRevA.82.043433}
  {\bibfield  {journal} {\bibinfo  {journal} {Phys. Rev. A}\ }\textbf {\bibinfo
  {volume} {82}},\ \bibinfo {pages} {043433} (\bibinfo {year}
  {2010})}\BibitemShut {NoStop}%
\bibitem [{\citenamefont {Livshits}\ and\ \citenamefont
  {Baer}(2006)}]{Livshits2006}%
  \BibitemOpen
  \bibfield  {author} {\bibinfo {author} {\bibfnamefont {E.}~\bibnamefont
  {Livshits}}\ and\ \bibinfo {author} {\bibfnamefont {R.}~\bibnamefont
  {Baer}},\ }\bibfield  {title} {\bibinfo {title} {Time-dependent
  density-functional studies of the d2 coulomb explosion},\ }\href
  {https://doi.org/10.1021/jp0600460} {\bibfield  {journal} {\bibinfo
  {journal} {The Journal of Physical Chemistry A}\ }\textbf {\bibinfo {volume}
  {110}},\ \bibinfo {pages} {8443} (\bibinfo {year} {2006})}\BibitemShut
  {NoStop}%
\bibitem [{\citenamefont {Runge}\ and\ \citenamefont
  {Gross}(1984)}]{runge1984density}%
  \BibitemOpen
  \bibfield  {author} {\bibinfo {author} {\bibfnamefont {E.}~\bibnamefont
  {Runge}}\ and\ \bibinfo {author} {\bibfnamefont {E.~K.}\ \bibnamefont
  {Gross}},\ }\bibfield  {title} {\bibinfo {title} {Density-functional theory
  for time-dependent systems},\ }\href@noop {} {\bibfield  {journal} {\bibinfo
  {journal} {Physical review letters}\ }\textbf {\bibinfo {volume} {52}},\
  \bibinfo {pages} {997} (\bibinfo {year} {1984})}\BibitemShut {NoStop}%
\bibitem [{\citenamefont {Ullrich}(2012)}]{ullrich}%
  \BibitemOpen
  \bibfield  {author} {\bibinfo {author} {\bibfnamefont {C.~A.}\ \bibnamefont
  {Ullrich}},\ }\href@noop {} {\emph {\bibinfo {title} {Time-Dependent
  Density-Functional Theory: Concepts and Applications}}}\ (\bibinfo
  {publisher} {Oxford University Press, USA},\ \bibinfo {year}
  {2012})\BibitemShut {NoStop}%
\bibitem [{\citenamefont {Kononov}\ \emph {et~al.}(2024)\citenamefont
  {Kononov}, \citenamefont {White}, \citenamefont {Nichols}, \citenamefont
  {Hu},\ and\ \citenamefont {Baczewski}}]{kononov_reproducibility_2024}%
  \BibitemOpen
  \bibfield  {author} {\bibinfo {author} {\bibfnamefont {A.}~\bibnamefont
  {Kononov}}, \bibinfo {author} {\bibfnamefont {A.~J.}\ \bibnamefont {White}},
  \bibinfo {author} {\bibfnamefont {K.~A.}\ \bibnamefont {Nichols}}, \bibinfo
  {author} {\bibfnamefont {S.~X.}\ \bibnamefont {Hu}},\ and\ \bibinfo {author}
  {\bibfnamefont {A.~D.}\ \bibnamefont {Baczewski}},\ }\bibfield  {title}
  {\bibinfo {title} {Reproducibility of real-time time-dependent density
  functional theory calculations of electronic stopping power in warm dense
  matter},\ }\href {https://doi.org/10.1063/5.0198008} {\bibfield  {journal}
  {\bibinfo  {journal} {Physics of Plasmas}\ }\textbf {\bibinfo {volume}
  {31}},\ \bibinfo {pages} {043904} (\bibinfo {year} {2024})}\BibitemShut
  {NoStop}%
\bibitem [{\citenamefont {Hekele}\ \emph {et~al.}(2021)\citenamefont {Hekele},
  \citenamefont {Yao}, \citenamefont {Kanai}, \citenamefont {Blum},\ and\
  \citenamefont {Kratzer}}]{hekele_all-electron_2021}%
  \BibitemOpen
  \bibfield  {author} {\bibinfo {author} {\bibfnamefont {J.}~\bibnamefont
  {Hekele}}, \bibinfo {author} {\bibfnamefont {Y.}~\bibnamefont {Yao}},
  \bibinfo {author} {\bibfnamefont {Y.}~\bibnamefont {Kanai}}, \bibinfo
  {author} {\bibfnamefont {V.}~\bibnamefont {Blum}},\ and\ \bibinfo {author}
  {\bibfnamefont {P.}~\bibnamefont {Kratzer}},\ }\bibfield  {title} {\bibinfo
  {title} {All-electron real-time and imaginary-time time-dependent density
  functional theory within a numeric atom-centered basis function framework},\
  }\href {https://doi.org/10.1063/5.0066753} {\bibfield  {journal} {\bibinfo
  {journal} {The Journal of Chemical Physics}\ }\textbf {\bibinfo {volume}
  {155}},\ \bibinfo {pages} {154801} (\bibinfo {year} {2021})}\BibitemShut
  {NoStop}%
\bibitem [{\citenamefont {Herring}\ and\ \citenamefont
  {Montemore}(2023)}]{herring_recent_2023}%
  \BibitemOpen
  \bibfield  {author} {\bibinfo {author} {\bibfnamefont {C.~J.}\ \bibnamefont
  {Herring}}\ and\ \bibinfo {author} {\bibfnamefont {M.~M.}\ \bibnamefont
  {Montemore}},\ }\bibfield  {title} {\bibinfo {title} {Recent {Advances} in
  {Real}-{Time} {Time}-{Dependent} {Density} {Functional} {Theory}
  {Simulations} of {Plasmonic} {Nanostructures} and {Plasmonic}
  {Photocatalysis}},\ }\href {https://doi.org/10.1021/acsnanoscienceau.2c00061}
  {\bibfield  {journal} {\bibinfo  {journal} {ACS Nanoscience Au}\ }\textbf
  {\bibinfo {volume} {3}},\ \bibinfo {pages} {269} (\bibinfo {year} {2023})},\
  \bibinfo {note} {publisher: American Chemical Society}\BibitemShut {NoStop}%
\bibitem [{\citenamefont {Xu}\ \emph {et~al.}(2024)\citenamefont {Xu},
  \citenamefont {Carney}, \citenamefont {Zhou}, \citenamefont {Shepard},\ and\
  \citenamefont {Kanai}}]{xu_real-time_2024}%
  \BibitemOpen
  \bibfield  {author} {\bibinfo {author} {\bibfnamefont {J.}~\bibnamefont
  {Xu}}, \bibinfo {author} {\bibfnamefont {T.~E.}\ \bibnamefont {Carney}},
  \bibinfo {author} {\bibfnamefont {R.}~\bibnamefont {Zhou}}, \bibinfo {author}
  {\bibfnamefont {C.}~\bibnamefont {Shepard}},\ and\ \bibinfo {author}
  {\bibfnamefont {Y.}~\bibnamefont {Kanai}},\ }\bibfield  {title} {\bibinfo
  {title} {Real-{Time} {Time}-{Dependent} {Density} {Functional} {Theory} for
  {Simulating} {Nonequilibrium} {Electron} {Dynamics}},\ }\href
  {https://doi.org/10.1021/jacs.3c08226} {\bibfield  {journal} {\bibinfo
  {journal} {Journal of the American Chemical Society}\ }\textbf {\bibinfo
  {volume} {146}},\ \bibinfo {pages} {5011} (\bibinfo {year} {2024})},\
  \bibinfo {note} {publisher: American Chemical Society}\BibitemShut {NoStop}%
\bibitem [{\citenamefont {Wang}(2015)}]{wang_efficient_2015}%
  \BibitemOpen
  \bibfield  {author} {\bibinfo {author} {\bibfnamefont {Z.}~\bibnamefont
  {Wang}},\ }\bibfield  {title} {\bibinfo {title} {Efficient {Real}-{Time}
  {Time}-{Dependent} {Density} {Functional} {Theory} {Method} and its
  {Application} to a {Collision} of an {Ion} with a {2D} {Material}},\
  }\bibfield  {journal} {\bibinfo  {journal} {Physical Review Letters}\
  }\textbf {\bibinfo {volume} {114}},\ \href
  {https://doi.org/10.1103/PhysRevLett.114.063004}
  {10.1103/PhysRevLett.114.063004} (\bibinfo {year} {2015})\BibitemShut
  {NoStop}%
\bibitem [{\citenamefont {Rossi}\ \emph {et~al.}(2017)\citenamefont {Rossi},
  \citenamefont {Kuisma}, \citenamefont {Puska}, \citenamefont {Nieminen},\
  and\ \citenamefont {Erhart}}]{rossi_kohnsham_2017}%
  \BibitemOpen
  \bibfield  {author} {\bibinfo {author} {\bibfnamefont {T.~P.}\ \bibnamefont
  {Rossi}}, \bibinfo {author} {\bibfnamefont {M.}~\bibnamefont {Kuisma}},
  \bibinfo {author} {\bibfnamefont {M.~J.}\ \bibnamefont {Puska}}, \bibinfo
  {author} {\bibfnamefont {R.~M.}\ \bibnamefont {Nieminen}},\ and\ \bibinfo
  {author} {\bibfnamefont {P.}~\bibnamefont {Erhart}},\ }\bibfield  {title}
  {\bibinfo {title} {Kohn{\textendash}{Sham} {Decomposition} in {Real}-{Time}
  {Time}-{Dependent} {Density}-{Functional} {Theory}: {An} {Efficient} {Tool}
  for {Analyzing} {Plasmonic} {Excitations}},\ }\href
  {https://doi.org/10.1021/acs.jctc.7b00589} {\bibfield  {journal} {\bibinfo
  {journal} {Journal of Chemical Theory and Computation}\ }\textbf {\bibinfo
  {volume} {13}},\ \bibinfo {pages} {4779} (\bibinfo {year} {2017})},\ \bibinfo
  {note} {publisher: American Chemical Society}\BibitemShut {NoStop}%
\bibitem [{\citenamefont {Noda}\ \emph {et~al.}(2019)\citenamefont {Noda},
  \citenamefont {Sato}, \citenamefont {Hirokawa}, \citenamefont {Uemoto},
  \citenamefont {Takeuchi}, \citenamefont {Yamada}, \citenamefont {Yamada},
  \citenamefont {Shinohara}, \citenamefont {Yamaguchi}, \citenamefont {Iida},
  \citenamefont {Floss}, \citenamefont {Otobe}, \citenamefont {Lee},
  \citenamefont {Ishimura}, \citenamefont {Boku}, \citenamefont {Bertsch},
  \citenamefont {Nobusada},\ and\ \citenamefont {Yabana}}]{noda_salmon_2019}%
  \BibitemOpen
  \bibfield  {author} {\bibinfo {author} {\bibfnamefont {M.}~\bibnamefont
  {Noda}}, \bibinfo {author} {\bibfnamefont {S.~A.}\ \bibnamefont {Sato}},
  \bibinfo {author} {\bibfnamefont {Y.}~\bibnamefont {Hirokawa}}, \bibinfo
  {author} {\bibfnamefont {M.}~\bibnamefont {Uemoto}}, \bibinfo {author}
  {\bibfnamefont {T.}~\bibnamefont {Takeuchi}}, \bibinfo {author}
  {\bibfnamefont {S.}~\bibnamefont {Yamada}}, \bibinfo {author} {\bibfnamefont
  {A.}~\bibnamefont {Yamada}}, \bibinfo {author} {\bibfnamefont
  {Y.}~\bibnamefont {Shinohara}}, \bibinfo {author} {\bibfnamefont
  {M.}~\bibnamefont {Yamaguchi}}, \bibinfo {author} {\bibfnamefont
  {K.}~\bibnamefont {Iida}}, \bibinfo {author} {\bibfnamefont {I.}~\bibnamefont
  {Floss}}, \bibinfo {author} {\bibfnamefont {T.}~\bibnamefont {Otobe}},
  \bibinfo {author} {\bibfnamefont {K.-M.}\ \bibnamefont {Lee}}, \bibinfo
  {author} {\bibfnamefont {K.}~\bibnamefont {Ishimura}}, \bibinfo {author}
  {\bibfnamefont {T.}~\bibnamefont {Boku}}, \bibinfo {author} {\bibfnamefont
  {G.~F.}\ \bibnamefont {Bertsch}}, \bibinfo {author} {\bibfnamefont
  {K.}~\bibnamefont {Nobusada}},\ and\ \bibinfo {author} {\bibfnamefont
  {K.}~\bibnamefont {Yabana}},\ }\bibfield  {title} {\bibinfo {title}
  {{SALMON}: {Scalable} {Ab}-initio {Light}{\textendash}{Matter} simulator for
  {Optics} and {Nanoscience}},\ }\href
  {https://doi.org/10.1016/j.cpc.2018.09.018} {\bibfield  {journal} {\bibinfo
  {journal} {Computer Physics Communications}\ }\textbf {\bibinfo {volume}
  {235}},\ \bibinfo {pages} {356} (\bibinfo {year} {2019})}\BibitemShut
  {NoStop}%
\bibitem [{\citenamefont {Andrade}\ \emph {et~al.}(2012)\citenamefont
  {Andrade}, \citenamefont {Alberdi-Rodriguez}, \citenamefont {Strubbe},
  \citenamefont {Oliveira}, \citenamefont {Nogueira}, \citenamefont {Castro},
  \citenamefont {Muguerza}, \citenamefont {Arruabarrena}, \citenamefont
  {Louie}, \citenamefont {Aspuru-Guzik}, \citenamefont {Rubio},\ and\
  \citenamefont {Marques}}]{andrade_time-dependent_2012}%
  \BibitemOpen
  \bibfield  {author} {\bibinfo {author} {\bibfnamefont {X.}~\bibnamefont
  {Andrade}}, \bibinfo {author} {\bibfnamefont {J.}~\bibnamefont
  {Alberdi-Rodriguez}}, \bibinfo {author} {\bibfnamefont {D.~A.}\ \bibnamefont
  {Strubbe}}, \bibinfo {author} {\bibfnamefont {M.~J.~T.}\ \bibnamefont
  {Oliveira}}, \bibinfo {author} {\bibfnamefont {F.}~\bibnamefont {Nogueira}},
  \bibinfo {author} {\bibfnamefont {A.}~\bibnamefont {Castro}}, \bibinfo
  {author} {\bibfnamefont {J.}~\bibnamefont {Muguerza}}, \bibinfo {author}
  {\bibfnamefont {A.}~\bibnamefont {Arruabarrena}}, \bibinfo {author}
  {\bibfnamefont {S.~G.}\ \bibnamefont {Louie}}, \bibinfo {author}
  {\bibfnamefont {A.}~\bibnamefont {Aspuru-Guzik}}, \bibinfo {author}
  {\bibfnamefont {A.}~\bibnamefont {Rubio}},\ and\ \bibinfo {author}
  {\bibfnamefont {M.~A.~L.}\ \bibnamefont {Marques}},\ }\bibfield  {title}
  {\bibinfo {title} {Time-dependent density-functional theory in massively
  parallel computer architectures: the octopus project},\ }\href
  {https://doi.org/10.1088/0953-8984/24/23/233202} {\bibfield  {journal}
  {\bibinfo  {journal} {Journal of Physics: Condensed Matter}\ }\textbf
  {\bibinfo {volume} {24}},\ \bibinfo {pages} {233202} (\bibinfo {year}
  {2012})}\BibitemShut {NoStop}%
\bibitem [{\citenamefont {Fuks}\ \emph {et~al.}(2016)\citenamefont {Fuks},
  \citenamefont {Nielsen}, \citenamefont {Ruggenthaler},\ and\ \citenamefont
  {Maitra}}]{fuks_time-dependent_2016}%
  \BibitemOpen
  \bibfield  {author} {\bibinfo {author} {\bibfnamefont {J.~I.}\ \bibnamefont
  {Fuks}}, \bibinfo {author} {\bibfnamefont {S.~E.~B.}\ \bibnamefont
  {Nielsen}}, \bibinfo {author} {\bibfnamefont {M.}~\bibnamefont
  {Ruggenthaler}},\ and\ \bibinfo {author} {\bibfnamefont {N.~T.}\ \bibnamefont
  {Maitra}},\ }\bibfield  {title} {\bibinfo {title} {Time-dependent density
  functional theory beyond kohn–sham slater determinants},\ }\href
  {https://doi.org/10.1039/C6CP00722H} {\bibfield  {journal} {\bibinfo
  {journal} {Phys. Chem. Chem. Phys.}\ }\textbf {\bibinfo {volume} {18}},\
  \bibinfo {pages} {20976} (\bibinfo {year} {2016})}\BibitemShut {NoStop}%
\bibitem [{\citenamefont {Dar}(2024)}]{dar_reformulation_2024}%
  \BibitemOpen
  \bibfield  {author} {\bibinfo {author} {\bibfnamefont {D.~B.}\ \bibnamefont
  {Dar}},\ }\bibfield  {title} {\bibinfo {title} {Reformulation of
  {Time}-{Dependent} {Density} {Functional} {Theory} for {Nonperturbative}
  {Dynamics}: {The} {Rabi} {Oscillation} {Problem} {Resolved}},\ }\bibfield
  {journal} {\bibinfo  {journal} {Physical Review Letters}\ }\textbf {\bibinfo
  {volume} {133}},\ \href {https://doi.org/10.1103/PhysRevLett.133.096401}
  {10.1103/PhysRevLett.133.096401} (\bibinfo {year} {2024})\BibitemShut
  {NoStop}%
\bibitem [{\citenamefont {Li}\ \emph {et~al.}(2020)\citenamefont {Li},
  \citenamefont {Govind}, \citenamefont {Isborn}, \citenamefont {DePrince},\
  and\ \citenamefont {Lopata}}]{doi:10.1021/acs.chemrev.0c00223}%
  \BibitemOpen
  \bibfield  {author} {\bibinfo {author} {\bibfnamefont {X.}~\bibnamefont
  {Li}}, \bibinfo {author} {\bibfnamefont {N.}~\bibnamefont {Govind}}, \bibinfo
  {author} {\bibfnamefont {C.}~\bibnamefont {Isborn}}, \bibinfo {author}
  {\bibfnamefont {A.~E.~I.}\ \bibnamefont {DePrince}},\ and\ \bibinfo {author}
  {\bibfnamefont {K.}~\bibnamefont {Lopata}},\ }\bibfield  {title} {\bibinfo
  {title} {Real-time time-dependent electronic structure theory},\ }\href
  {https://doi.org/10.1021/acs.chemrev.0c00223} {\bibfield  {journal} {\bibinfo
   {journal} {Chemical Reviews}\ }\textbf {\bibinfo {volume} {120}},\ \bibinfo
  {pages} {9951} (\bibinfo {year} {2020})},\ \bibinfo {note} {pMID:
  32813506}\BibitemShut {NoStop}%
\bibitem [{\citenamefont {Varga}\ and\ \citenamefont
  {Driscoll}(2011)}]{Varga_Driscoll_2011a}%
  \BibitemOpen
  \bibfield  {author} {\bibinfo {author} {\bibfnamefont {K.}~\bibnamefont
  {Varga}}\ and\ \bibinfo {author} {\bibfnamefont {J.~A.}\ \bibnamefont
  {Driscoll}},\ }in\ \href@noop {} {\emph {\bibinfo {booktitle} {Computational
  Nanoscience: Applications for Molecules, Clusters, and Solids}}}\ (\bibinfo
  {publisher} {Cambridge University Press},\ \bibinfo {year}
  {2011})\BibitemShut {NoStop}%
\bibitem [{\citenamefont {Troullier}\ and\ \citenamefont
  {Martins}(1991)}]{PhysRevB.43.1993}%
  \BibitemOpen
  \bibfield  {author} {\bibinfo {author} {\bibfnamefont {N.}~\bibnamefont
  {Troullier}}\ and\ \bibinfo {author} {\bibfnamefont {J.~L.}\ \bibnamefont
  {Martins}},\ }\bibfield  {title} {\bibinfo {title} {Efficient
  pseudopotentials for plane-wave calculations},\ }\href
  {https://doi.org/10.1103/PhysRevB.43.1993} {\bibfield  {journal} {\bibinfo
  {journal} {Phys. Rev. B}\ }\textbf {\bibinfo {volume} {43}},\ \bibinfo
  {pages} {1993} (\bibinfo {year} {1991})}\BibitemShut {NoStop}%
\bibitem [{\citenamefont {Perdew}\ and\ \citenamefont
  {Zunger}(1981)}]{PhysRevB.23.5048}%
  \BibitemOpen
  \bibfield  {author} {\bibinfo {author} {\bibfnamefont {J.~P.}\ \bibnamefont
  {Perdew}}\ and\ \bibinfo {author} {\bibfnamefont {A.}~\bibnamefont
  {Zunger}},\ }\bibfield  {title} {\bibinfo {title} {Self-interaction
  correction to density-functional approximations for many-electron systems},\
  }\href {https://doi.org/10.1103/PhysRevB.23.5048} {\bibfield  {journal}
  {\bibinfo  {journal} {Phys. Rev. B}\ }\textbf {\bibinfo {volume} {23}},\
  \bibinfo {pages} {5048} (\bibinfo {year} {1981})}\BibitemShut {NoStop}%
\bibitem [{\citenamefont {Manolopoulos}(2002)}]{10.1063/1.1517042}%
  \BibitemOpen
  \bibfield  {author} {\bibinfo {author} {\bibfnamefont {D.~E.}\ \bibnamefont
  {Manolopoulos}},\ }\bibfield  {title} {\bibinfo {title} {{Derivation and
  reflection properties of a transmission-free absorbing potential}},\ }\href
  {https://doi.org/10.1063/1.1517042} {\bibfield  {journal} {\bibinfo
  {journal} {The Journal of Chemical Physics}\ }\textbf {\bibinfo {volume}
  {117}},\ \bibinfo {pages} {9552} (\bibinfo {year} {2002})},\ \Eprint
  {https://arxiv.org/abs/https://pubs.aip.org/aip/jcp/article-pdf/117/21/9552/19225128/9552\_1\_online.pdf}
  {https://pubs.aip.org/aip/jcp/article-pdf/117/21/9552/19225128/9552\_1\_online.pdf}
  \BibitemShut {NoStop}%
\bibitem [{\citenamefont {Boerner}\ \emph {et~al.}(2023)\citenamefont
  {Boerner}, \citenamefont {Deems}, \citenamefont {Furlani}, \citenamefont
  {Knuth},\ and\ \citenamefont {Towns}}]{aces}%
  \BibitemOpen
  \bibfield  {author} {\bibinfo {author} {\bibfnamefont {T.~J.}\ \bibnamefont
  {Boerner}}, \bibinfo {author} {\bibfnamefont {S.}~\bibnamefont {Deems}},
  \bibinfo {author} {\bibfnamefont {T.~R.}\ \bibnamefont {Furlani}}, \bibinfo
  {author} {\bibfnamefont {S.~L.}\ \bibnamefont {Knuth}},\ and\ \bibinfo
  {author} {\bibfnamefont {J.}~\bibnamefont {Towns}},\ }\bibfield  {title}
  {\bibinfo {title} {Access: Advancing innovation: Nsf’s advanced
  cyberinfrastructure coordination ecosystem: Services \& support},\ }in\ \href
  {https://doi.org/10.1145/3569951.3597559} {\emph {\bibinfo {booktitle}
  {Practice and Experience in Advanced Research Computing 2023: Computing for
  the Common Good}}},\ \bibinfo {series and number} {PEARC '23}\ (\bibinfo
  {publisher} {Association for Computing Machinery},\ \bibinfo {address} {New
  York, NY, USA},\ \bibinfo {year} {2023})\ pp.\ \bibinfo {pages}
  {173--176}\BibitemShut {NoStop}%
\end{thebibliography}
%

\end{document}